\documentstyle[aps,multicol,epsf]{revtex}

\begin{document}
\draft

\title{
 Hall Effect and Resistivity in High-$T_{\rm c}$ Superconductors: \\
 The Conserving Approximation}

\author{Hiroshi Kontani, Kazuki Kanki$^{1}$, and Kazuo Ueda}

\address{
Institute for Solid State Physics, University of Tokyo, 
7-22-1 Roppongi, Minato-ku, Tokyo 106-8666 \\
$^{1}$College of Integrated Arts and Sciences, Osaka 
Prefecture University, Sakai 599-8531
}

\date{October 15, 1998}
\maketitle

\newcommand{\simle}
{\raisebox{-0.75ex}[-1.5ex]{$\;\stackrel{<}{\sim}\;$}}
\newcommand{\simge}
{\raisebox{-0.75ex}[-1.5ex]{$\;\stackrel{>}{\sim}\;$}}
\def\d{{\partial}}
\def\s{{\sigma}}
\def\e{{\epsilon}}
\def\k{{ {\bf k} }}
\def\p{{ {\bf p} }}
\def\q{{ {\bf q} }}
\def\Q{{ {\bf Q} }}
\def\x{{ {\bf x} }}
\def\w{{\omega}}
\def\a{{\alpha}}
\def\b{{\beta}}
\def\v{{\varphi}}
\def\g{{\gamma}}
\def\G{{\Gamma}}        
\def\l{{\lambda}}
\def\L{{\Lambda}}       
\def\D{{\Delta}}
\def\i{{ {\rm i} }}
\def\expo{{ {\rm e} }}
\def\I{{ \mbox{\scriptsize I} }}
\def\II{{ \mbox{\scriptsize II} }}
\def\III{{ \mbox{\scriptsize III} }}
\def\IV{{ \mbox{\scriptsize IV} }}

\begin{abstract}
The Hall coefficient, $R_{\rm H}$, of high-$T_{\rm c}$ cuprates
in the normal state shows the striking non-Fermi liquid behavior:
$R_{\rm H}$ follows a Curie-Weiss type temperature dependence,
and $|R_{\rm H}| \gg 1/|n{e}|$ 
at low temperatures in the under-doped compounds.
Moreover, $R_{\rm H}$ is positive for hole-doped compounds
and is negative for electron-doped ones,
although each of them has a similar hole-like Fermi surface.
In this paper, we give the explanation of this long-standing problem
from the standpoint of the 
nearly antiferromagnetic (AF) Fermi liquid.
We consider seriously the vertex corrections for the current
which are indispensable to satisfy the conservation laws,
which are violated within the conventional Boltzmann 
transport approximation.
The obtained total current ${\vec J}_\k$ takes an enhanced value and
is no more perpendicular to the Fermi surface
due to the strong AF fluctuations.
By virtue of this mechanism, 
the anomalous behavior of $R_{\rm H}$
in high-T$_{\rm c}$ cuprates is neutrally explained.
We find that both the temperature and the (electron, or hole) 
doping dependences of $R_{\rm H}$ in high-$T_{\rm c}$ cuprates 
are reproduced well by numerical calculations
based on the fluctuation-exchange (FLEX) approximation,
applied to the single-band Hubbard model.
We also discuss the temperature dependence of 
$R_{\rm H}$ in other nearly AF metals,
e.g.,
V$_2$O$_3$, $\kappa$-BEDT-TTF organic superconductors,
and heavy fermion systems close to the AF phase boundary.
\end{abstract}



\begin{multicols}{2}
\narrowtext

\section{Introduction}

In the normal state of high-$T_{\rm c}$ superconductors (HTSC's), 
various quantities deviate from the conventional Fermi liquid behaviors.
 \cite{review-Iye}
These non-Fermi liquid features have been studied intensively 
both theoretically and experimentally
because they have close relation to the 
mechanism of superconductivity.
For example, the electrical resistivity ($\rho$) and 
the longitudinal NMR relaxation rate ($1/T_1$) in HTSC 
show universally the behaviors $\rho \propto T$, $1/T_1 \propto T^0$
for a wide range of temperatures.
 \cite{review-Asayama}
These are quite different from the conventional Fermi liquid behaviors,
$\rho \propto T^2$, $1/T_1 \propto T$.

In HTSC's, the Hall coefficient ($R_{\rm H}$) also shows an
interesting non-Fermi liquid behavior:
It shows a drastic temperature dependence 
although the Fermi surface (FS) in HTSC is non-degenerate 
and its shape is simple.
At high temperatures ($\sim1000$K), $R_{\rm H}$ takes 
a nearly constant value, and its doping dependence is very small.
Its value is close to the one estimated by
the LDA band calculation, $R_{\rm H}^{\rm band}$.
 \cite{Hall-band}
The doping dependence of $R_{\rm H}$ is also very small there.

\begin{figure}
\epsfxsize=55mm
\centerline{\epsffile{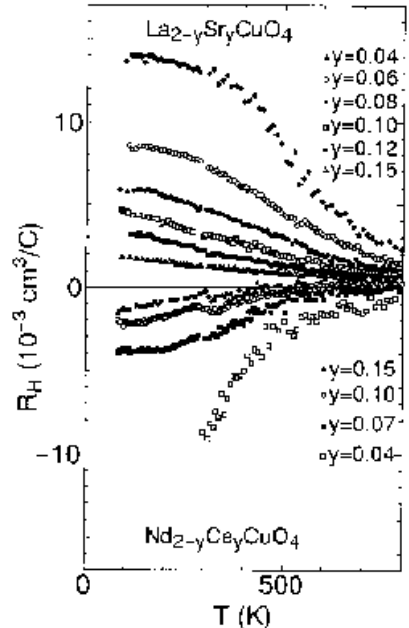}}
\caption{
 Temperature dependence of $R_{\rm H}$ in LSCO (hole-doping) 
 and NCCO (electron-doping) in the paramagnetic state.
 Note that $1/|ne|\sim 1.5\times10^{-3}{\rm cm}^3$C in HTSC's.}
\label{fig:Sato}
\end{figure}

On the other hand, as the temperature decreases, $R_{\rm H}$
begins to show a Curie-Weiss type temperature dependence, and
its maximum value is a few times larger than $R_{\rm H}^{\rm band}$
at the optimum-doping.
This enhancement of $R_{\rm H}$ further increases
in the under-doped region.
In the hole-doped compounds,
e.g., YBa$_2$Cu$_3$O$_{7-\delta}$ (YBCO) or 
La$_{2-\delta}$Sr$_\delta$CuO$_4$ (LSCO),
$dR_{\rm H}/dT<0$ is observed and $R_{\rm H}$ is positive
for a wide range of temperatures.
 \cite{Sato,Takagi,Nishikawa,Peng,Chien,Kubo}
On the other hand, $dR_{\rm H}/dT>0$ is realized
in the electron-doped compounds,
e.g., Nd$_{2-\delta}$Ce$_\delta$CuO$_4$ (NCCO), 
so the sign of $R_{\rm H}$ 
changes to negative at a low temperature
although its FS is hole-like.
 \cite{Sato,Uchida,Fournier}
Figure \ref{fig:Sato} shows a summary of experimental results of
LSCO and NCCO in the under-doped region,
where an approximate electron-hole symmetry is realized.
 \cite{Sato}
In both compounds, $|R_{\rm H}|$ increases near the half-filling.

The Hall effect is one of the unsolved problems in HTSC.
Its unusual features mentioned above 
are summarized as follows:
(i) The Curie-Weiss type behavior of $R_{\rm H}$ 
 in a quite wide range of temperatures.
(ii) The enhancement of $R_{\rm H}$ in the under-doped region.
(iii) $R_{\rm H}<0$ in the electron-doped compounds.

Nowadays,
various non-Fermi liquid phenomena of HTSC
have been explained by using different types of
spin-fluctuation theories, e.g.,
the SCR theory,
 \cite{SCR}
the spin-fluctuation model,
 \cite{Monthoux-Pines}
and the FLEX theory.
 \cite{Bickers-White,Monthoux-Scalapino}
They can explain a reasonable $T_{\rm c}$ of the 
$d_{x^2-y^2}$ superconductivity.
They can also explain the pseudo gap formation 
in the density of states,
 \cite{shadow-band,incipient-AF,Dahm-Tewordt}
the shadow band formation, 
 \cite{shadow-band}
and the collective modes emerging below $T_{\rm c}$.
 \cite{Morr,Takimoto}

So far, various attempts have been made 
on the Hall effect in HTSC.
 \cite{Miyake,Imada,Hall-Pines,Yanase}
Some of them are in the framework of the spin fluctuation model,
by using the Boltzmann transport approximation
(or one-loop approximation).
 \cite{Hall-Pines,Yanase}
However, the Boltzmann approximation can not explain the 
magnitude of $|R_{\rm H}|(\gg 1/ne)$ in the under-doped region.
Moreover, it predicts $R_{\rm H}>0$ for both YBCO and NCCO
because they have similar hole-like FS's.
This result contradicts with the experiments, shown in Fig. \ref{fig:Sato}.
Therefore, the behavior of $R_{\rm H}$ noted as (i)-(iii) above
have casted some suspicion on the validity of the 
nearly AF Fermi liquid description for the HTSC.

In this paper, we study the Hall effect of HTSC's
based on a conserving approximation.
 \cite{Baym-Kadanoff,Baym}
We use the expressions for the conductivity 
and the Hall conductivity 
derived from the Kubo formula.
 \cite{Eliashberg,Yamada-Yosida,Kohno-Yamada,Fukuyama}
Then, we study the vertex corrections for the current
according to the conserving approximation,
and find out that they show a 
critical behavior as a natural consequence of the strong 
backward scatterings by the AF fluctuations.
By virtue of this fact,
we succeed in explaining the overall features of 
$R_{\rm H}$, noted as (i)-(iii) above, 
without assuming a non-Fermi liquid ground state.
We also find that the conventional Boltzmann approximation,
where the conservation laws are violated,
cannot reproduce any of (i)-(iii).

We analyze the extended Hubbard model
as an effective model for HTSC's.
We use the fluctuation exchange (FLEX) approximation to 
calculate the Green function and the self-energy.
 \cite{Bickers-White,Monthoux-Scalapino}
It is a kind of self-consistent perturbation theory with respect to $U$,
and it has advantages for handling large spin fluctuations.

Phenomenologically, the spin propagator in HTSC's
is expressed for small $\q$ and $\w$ as follows:
 \cite{SCR,Monthoux-Pines,Hall-Pines}
\begin{eqnarray}
 \chi_{\q}^{s}(\w) = \frac{\chi_Q}{1+\xi^2 (\q-\Q)^2 + \i \w/\w_{\rm sf}},
 \label{eqn:kai_qw}
\end{eqnarray}
where $\Q$ is the antiferromagnetic (AF) wavevector, and $\xi$
is the AF correlation length.
Experimentally, $\xi^2$ follows a Curie-Weiss type temperature 
dependence at higher temperatures.
It ceases to increase at $T_{\rm c}$ in the over-doped region,
or at the characteristic temperature $T^\ast (>T_{\rm c})$
in the under-doped region.
We call $T^\ast$ the pseudo spin-gap temperature as usual.
In general, the following relations are satisfied for $T>T^\ast$ 
experimentally:
 \cite{Mag_scaling}
\begin{eqnarray}
& &\xi^2 \approx \a_0/( \ T+\Theta \ ), 
 \label{eqn:parameters1} \\
& &\chi_Q \approx \a_1\cdot \xi^2, \ \ \
 1/\w_{\rm sf} \approx \a_2\cdot \xi^2,
 \label{eqn:parameters2}
\end{eqnarray}
where $\Theta$, $\a_0$, $\a_1$ and $\a_2$ are constants.
 \cite{pseudo_scaling}
The coefficient $\a_0$ increases rapidly in the under-doped region, 
and $\xi^2$ reaches $\sim O(100)$ at $T^\ast$ 
nearby the half-filling.
(We put the unit-cell length $a=1$.)
And $\a_2$ decreases rather moderately in the under-doped region.
The relation $\w_{\rm sf} \simge T$ ($\w_{\rm sf} \simle T$)
is satisfied in the over-doped (under-doped) region.

The typical spin-fluctuation theories reproduce the 
experimental relations
(\ref{eqn:parameters1}) and (\ref{eqn:parameters2})
for $T>T^\ast$.
 \cite{Dahm-Tewordt}
Moreover, the approximate relations
$\rho  \propto \xi^2 T^2$,
$1/T_1 \propto \xi^2 T$
are derived in the nearly AF Fermi liquid.
 \cite{SCR,Hall-Pines,Yanase,Kohno-Yamada2}
So, the experimental non-Fermi liquid behaviors of 
$\rho$ and $1/T_1$ are naturally explained by the spin-fluctuation theories.
One may also expect that the anomalous behavior of 
$R_{\rm H}$ in HTSC comes from the $T$-dependence of $\xi$.
In this paper, we find that $R_{\rm H}\propto\xi^2$ is realized
through the vertex corrections for the current.
A similar study 
based on the phenomenological AF spin-fluctuation model 
is reported in another paper.
 \cite{Kanki}

The contents of this paper are as follows:
In \S II, we introduce the single-band Hubbard model
with some sets of parameters 
corresponding to YBCO, LSCO and NCCO.
In \S III, we review the general formulation for $\s_{xx}$
and $\s_{xy}/H$ based on the Fermi liquid theory,
and rewrite $\s_{xy}/H$ into a simpler form.
In \S IV, the vertex corrections to the current is studied
by using the conserving approximation.
We find that only the Maki-Thompson term is dominant.
In \S V, we solve the Bethe-Salpeter equation for the total
current ${\vec J}_\k$ analytically,
and the relation $R_{\rm H}\propto \xi^2$ is derived.
In \S VI, numerical results for $\rho$ and $R_{\rm H}$ obtained
by the FLEX theory are presented, which are consistent
with the experimental behaviors in HTSC's.
Finally, in \S VII, 
the Hall effect in heavy fermion systems are discussed briefly.

The readers who are mainly interested in the numerical calculation
of $R_{\rm H}$ can proceed to \S VI.B for the first reading,
where the sufficient set of equations for the numerical calculations
for $\s_{xx}$ and $\s_{xy}/H$ are explained shortly.

\section{Model Hamiltonian}
In this paper, we treat the following extended Hubbard model
with ($U,t_0,t_1,t_2$):
\begin{eqnarray}
& &H = \sum_{\k\s} \e_\k^0 c_{\k\s}^\dagger c_{\k\s} + 
  U\sum_{\k\k'\q} c_{\k+\q\uparrow}^\dagger c_{\k'-\q\downarrow}^\dagger 
  c_{\k'\downarrow} c_{\k\uparrow}, \\
& &\e_\k^0= 2t_0(\cos(k_x)+\cos(k_y)) + 4t_1\cos(k_x)\cos(k_y) 
  \nonumber \\
& &\ \ \ \   + 2t_2(\cos(2k_x)+\cos(2k_y)),
\end{eqnarray}
where $c_{\k\s}^\dagger$ is the creation operator of an electron
with momentum $\k$ and spin $\s$, and $U$ is the on-site Coulomb
repulsion.
We represent the filling of the electrons by $n$,
and $n=1$ corresponds to the half-filling.

\begin{figure}
\epsfxsize=70mm
\centerline{\epsffile{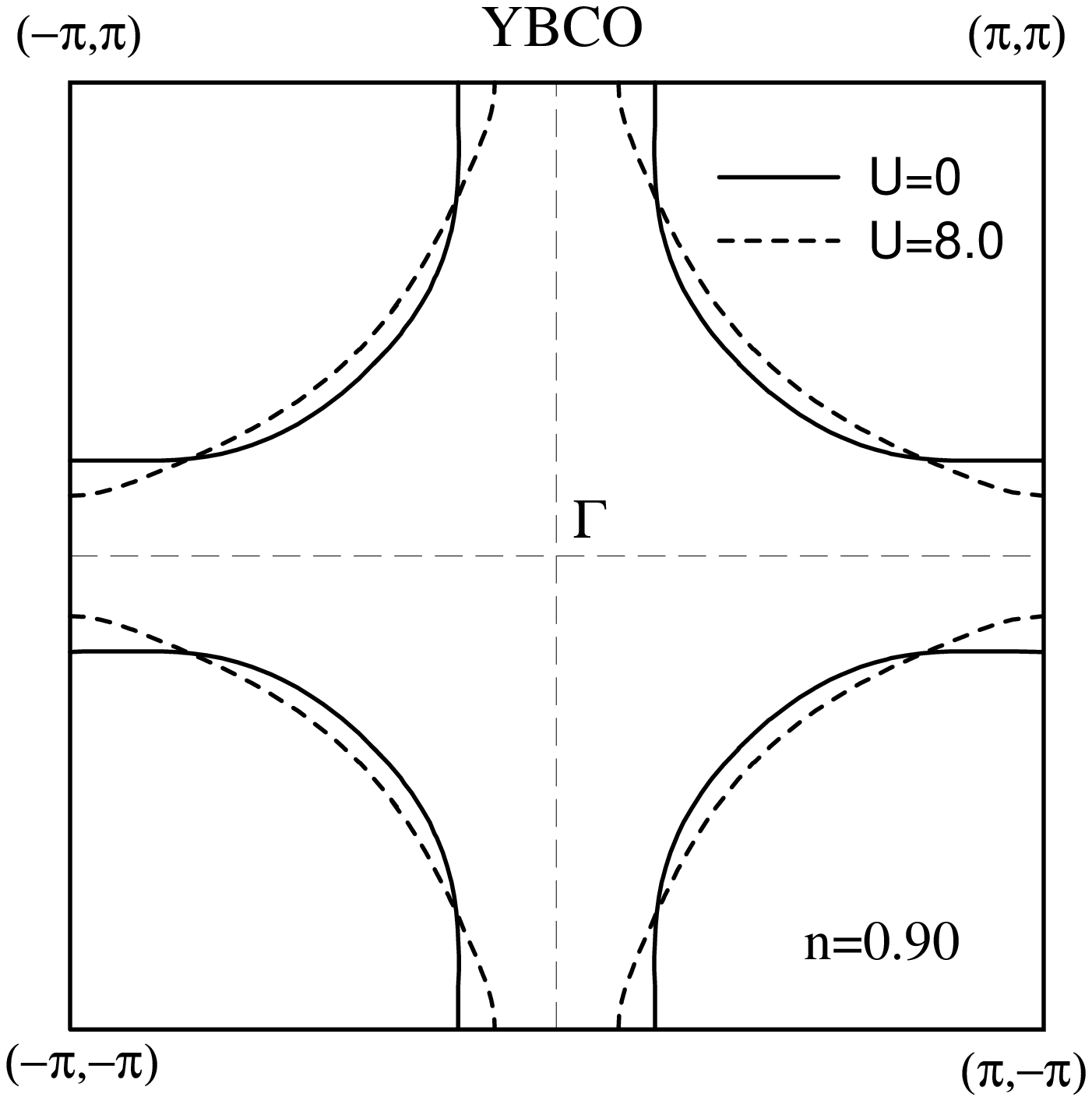}}
\epsfxsize=70mm
\centerline{\epsffile{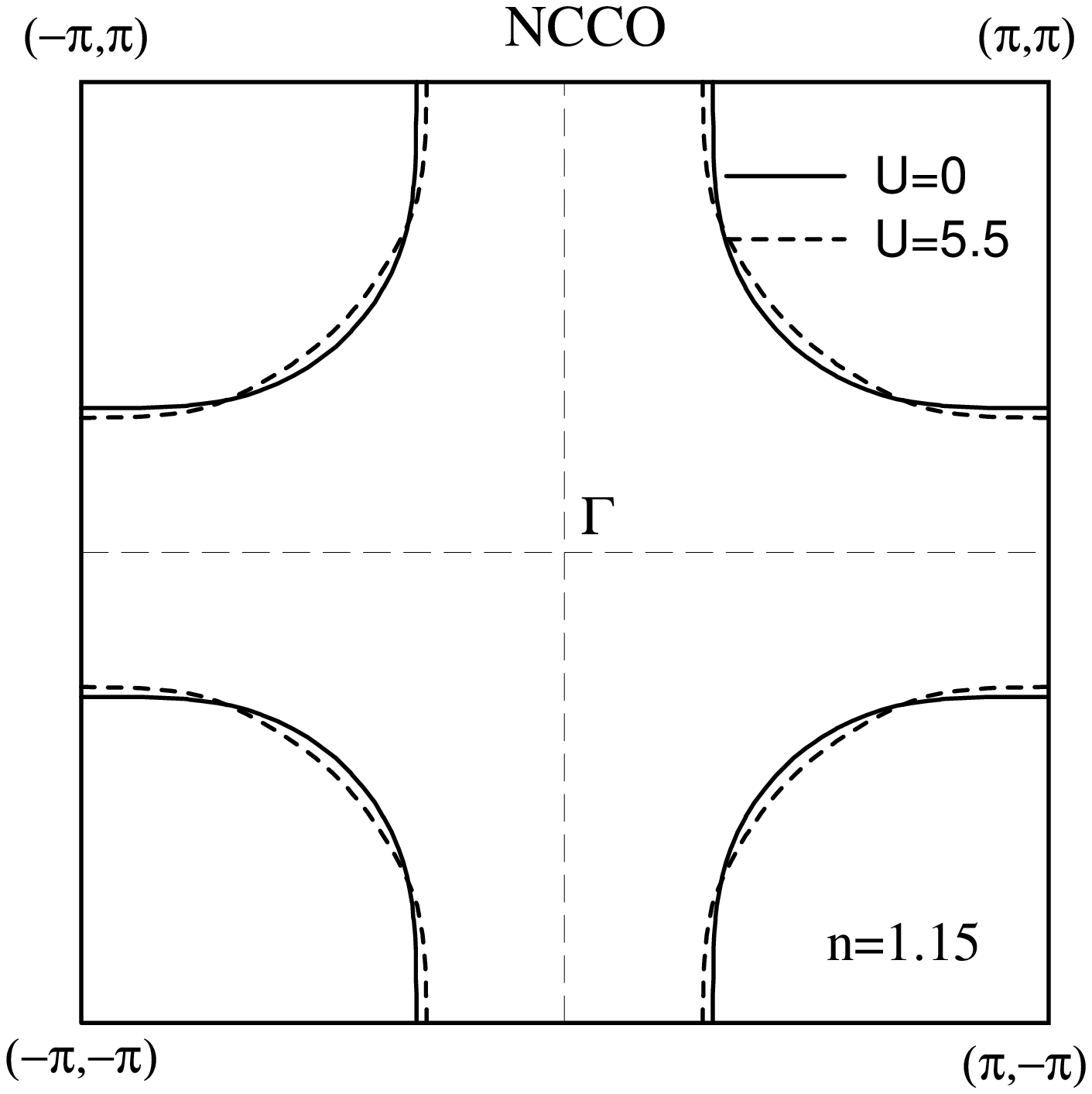}}
\epsfxsize=70mm
\centerline{\epsffile{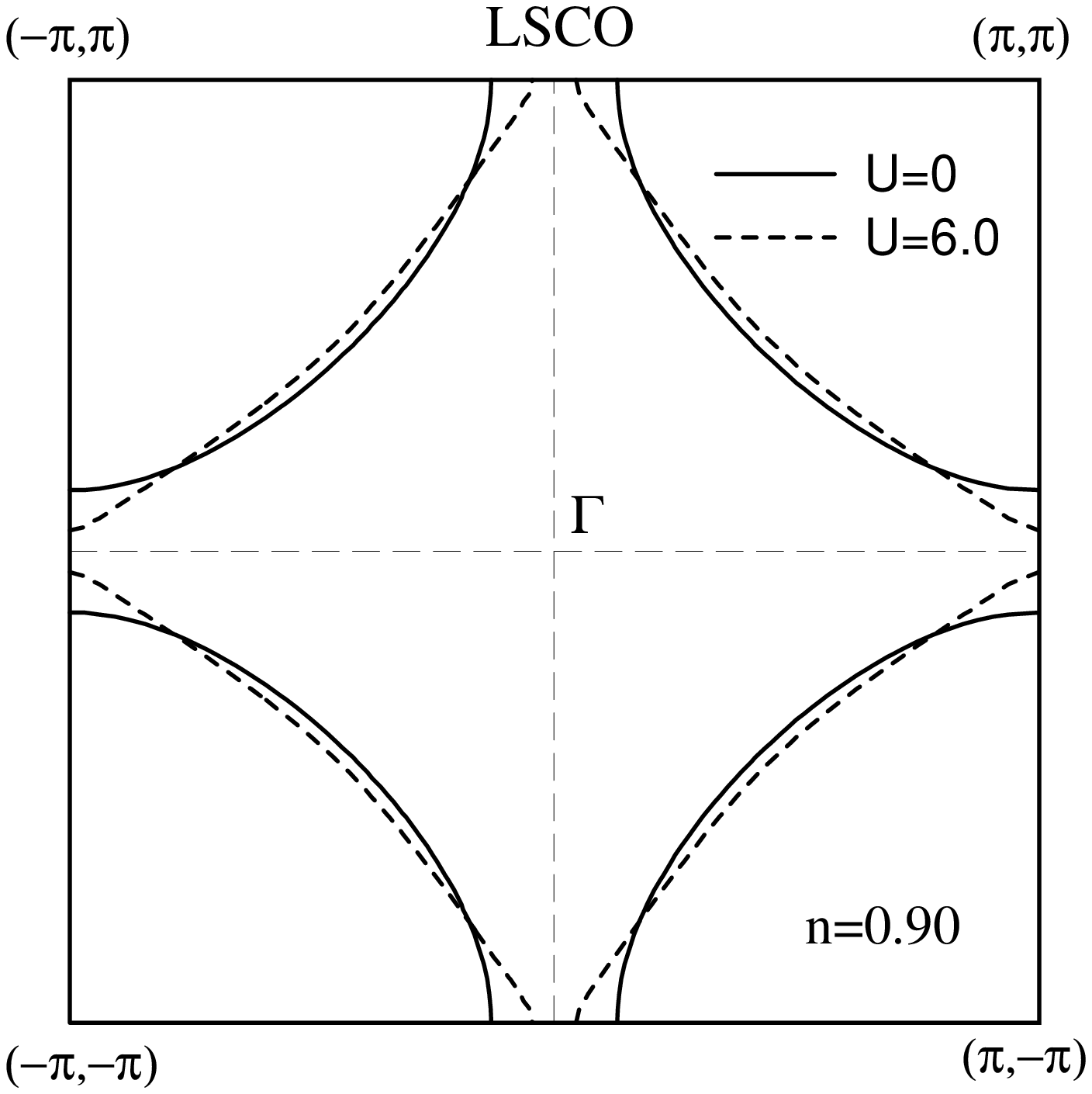}}
\caption{Fermi surface of (a)YBCO, (b)NCCO, and (c)LSCO.
 In (a) and (b), $t_0=-1$, $t_1=1/6$, and  $t_2=-1/5$.
 In (c), $t_0=-1$, $t_1=1/10$, and $t_2=-1/10$.}
\label{fig:FS-U}
\end{figure}

Taking the results by the LDA band calculation into account,
 \cite{YBCO-band,NCCO-band,LSCO-band,LSCO-band2}
we choose the following set of parameters:
(I) YBCO (hole-doping), NCCO (electron-doping):
 $t_0=-1$, $t_1=1/6$, $t_2=-1/5$.
 \cite{YBCO-band,NCCO-band,Tanamoto}
(II) LSCO (hole-doping):
  $t_0=-1$, $t_1=1/10$, $t_2=-1/10$.
 \cite{LSCO-band,LSCO-band2}
Figure \ref{fig:FS-U} shows the Fermi surfaces (FS's) for $U=0$
together with those for finite $U$ calculated by 
the FLEX approximation at $T=0.02$.
In the case of (I), 
the spectrum at $(\pi,0)$ is below the chemical potential $\mu$ 
at least for $n>0.6$, 
and the FS is hole-like everywhere.
On the other hand, in the case of (II)
the spectrum at $(\pi,0)$ is above $\mu$ 
for $n<0.77$ at $U=0$, and for $n\simle0.85$ at $U=6$.
These characters of the FS's 
coincide qualitatively with those by the LDA band calculations
 \cite{YBCO-band,NCCO-band,LSCO-band,LSCO-band2}
or by the angle resolved photoemission (ARPES) experiments.
 \cite{YBCO-ARPES,Shen,flat-band}

\begin{figure}
\epsfxsize=50mm
\centerline{\epsffile{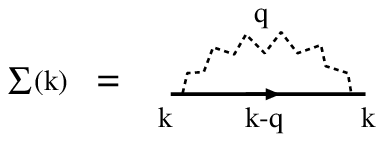}}
\caption{The self-energy of the FLEX theory.
 The full line and the wavy line represent $G(k-q)$ and $V(q)$,
 respectively.}
\label{fig:self-energy}
\end{figure}

Here, we summarize the formalism of the FLEX theory which will be
used throughout this paper.
The Dyson equation is written as
\begin{eqnarray}
\left\{ G_\k(\e_n) \right\}^{-1}
= \i\e_n+\mu-\e_\k^0 - \Sigma_\k(\e_n).
 \label{eqn:Dyson}
\end{eqnarray}
The self-energy $\Sigma_\k(\e_n)$ given by the FLEX approximation is
\begin{eqnarray}
& &\Sigma_{\k}(\e_n) 
 = T\sum_{\q,l} G_{\k-\q}(\e_n-\w_l)\cdot V_\q(\w_l),
 \label{eqn:self} \\
& &V_\q(\w_l)
= U^2 \left(\frac32 {\chi}_{\q}^{s}(\w_l) +\frac12 {\chi}_{\q}^{c}(\w_l) 
  - {\chi}_{\q}^0(\w_l) \right) +U \mbox{,}
     \label{eqn:def_V} \\
& &{\chi}_{\q}^{s}(\w_l)
 = {\chi}_{\q}^0 \cdot \left\{ 1- 
 U{\chi}_{\q}^0(\w_l) \right\}^{-1} \mbox{,} \\
     \label{eqn:chi_s}
& &{\chi}_{\q}^{c}(\w_l)
 = {\chi}_{\q}^0 \cdot \left\{ 1+
  U{\chi}_{\q}^0(\w_l) \right\}^{-1} \mbox{,}
     \label{eqn:chi_c} \\
& &\chi_{\q}^0(\w_l)
 = -T\sum_{\k, n} G_{\q+\k}(\w_l+\e_n) G_{\k}(\e_n) \mbox{,}
     \label{eqn:chi0}
\end{eqnarray}
where $\e_n= (2n+1)\pi T$ and $\w_l= 2l\pi T$, respectively.
The self-energy is shown by Fig. \ref{fig:self-energy}.
We solve the equations (\ref{eqn:Dyson})-(\ref{eqn:chi0}) self-consistently,
choosing $\mu$ so as to keep the filling constant,
$n=T\sum_{\k,n}G_\k(\e_n)\cdot{\rm e}^{-\i\e_n\cdot\delta}$,
where $\delta= 0^+$.

In a Fermi liquid, the real-frequency Green function in the vicinity
of $\w\sim0$ and $|\k|\sim k_{\rm F}$ is represented as
\begin{eqnarray}
G_\k(\w)= z_\k/(\w+\mu-\e_\k-\i z_\k\Delta_\k),
\end{eqnarray}
where $z_\k$ is the renormalization factor given by
$z_\k= (1-\d{\rm Re}\Sigma_\k(\w)/\d\w)^{-1}$,
$\e_\k$ is the quasi-particle spectrum given by the solution of
$\{{\rm Re}G_\k(\w)\}^{-1}=0$,
and $\Delta_\k=-{\rm Im}\Sigma_\k(\w+i\delta)>0$.
The density of states (DOS) is given by
\begin{eqnarray}
\rho_\k(\w)= -\frac1\pi {\rm Im}G_\k(\w+\i\delta).
\end{eqnarray}
In the case of $z_\k \Delta_\k\ll T$, 
$\rho_\k(\w)= z_\k\cdot \delta(\w+\mu-\e_\k)$.

The FLEX approximation is suitable for the analysis of the 
nearly AF Fermi liquid.
It has been applied to the square lattice Hubbard model
by many authors.
 \cite{Bickers-White,shadow-band,incipient-AF,Dahm-Tewordt,Takimoto,d-p-model}
Though it is an approximation, imaginary time Green function 
obtained by the FLEX agrees with the results by the QMC simulations
very well for a moderate $U$.
 \cite{Bickers-White}
Recently, it has also been  applied to the superconducting 
ladder compound, Sr$_{14-x}$Ca$_x$Cu$_{24}$O$_{41}$,
 \cite{Trellis}
and the organic superconducting $\kappa$-(BEDT-TTF) compounds.
 \cite{Kino,Kondo}

\begin{figure}
\epsfxsize=70mm
\centerline{\epsffile{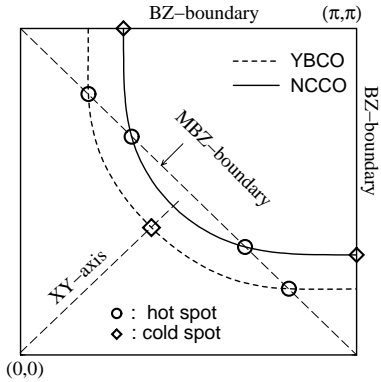}}
\caption{ The hot spots and the cold spots in
 YBCO and NCCO, respectively.}
\label{fig:FS-hotcold}
\end{figure}

We comment on the anisotropy of $\Delta_\k$ on the FS,
which becomes larger as the AF fluctuations grows at low temperatures.
$\Delta_\k$ takes a large value around the crossing points with
the magnetic Brillouin zone (MBZ)-boundary,
which we call {\it hot spots} as often referred to in literatures.
 \cite{Hall-Pines,Yanase}
And $\Delta_\k$ becomes small at the points 
where the distance from the MBZ-boundary is the largest,
which are called {\it cold spots}.
(see Fig. \ref{fig:FS-hotcold}.)
These cold spots play the major role for $\rho$ and $R_{\rm H}$.
We study this subject in \S VI in detail.

Finally, we discuss the validity and the limitation
of the FLEX theory on HTSC's.
In the FLEX approximation,
eqs. (\ref{eqn:parameters1}) and (\ref{eqn:parameters2}) 
are satisfied well, and the coefficient
$\a_0$ in eq. (\ref{eqn:parameters1})
increases rapidly as $n$ approaches to the half-filling ($n=1$). 
Moreover, the relation $\w_{\rm sf}\simle T (\simge T)$ in the 
under-doped (over-doped) region is satisfied qualitatively
as shown in Table I, 
which is consistent with experiments.
However, the FLEX approximation can not reproduce the 
experimentally observed pseudo spin-gap behaviors
for $T^{\ast}>T>T_{\rm c}$,
where $\xi$ ceases to increase
and $1/\w_{\rm sf}$ begins to decrease as the temperature decreases.
It would also be inapplicable near the Mott-insulating state,
i.e., $0.9\simge n \simge 1.1$.
In this paper we perform numerical studies for 
$n\le0.9$ or $n\ge1.1$, where the FLEX approximation gives 
reasonable results.

\vskip5mm\noindent
Table I \ : \ 
The value of $\w_{\rm sf}$ for $n=0.90$(under-doped), 
$n=0.85$(nearly optimum), and $n=0.80$(over-doped) obtained 
by the FLEX approximation at $T=0.02$.
\begin{center}
\begin{tabular}{cccc} \hline\hline
      & $n=0.90$ \ \ & $n=0.85$ \ \ & $n=0.80$ \\ \hline
 YBCO ($U\!=\!8$) \ \ & 0.018    & 0.034    & 0.046    \\ \hline
 LSCO ($U\!=\!6$) \ \ & 0.013    & 0.019    & 0.024    \\ \hline\hline
\end{tabular}
\end{center}
\vskip5mm

\section{Formalism of conductivity in the Fermi liquid theory}
In this section, we review the transport theory.
By the Kubo formula, the conductivity is given by 
\begin{eqnarray}
& &\s_{\mu\nu}= {e}^2 \sum_{\k\k'\s\s'} v_{\k\mu}^0 v_{\k'\nu}^0
 \cdot \left.\frac{ {\rm Im} K_{\k\s,\k'\s'}(\w+i\delta) }{\w}
  \right|_{\w=0},
 \label{eqn:Kubo}
 \\
& &K_{\k\s,\k'\s'}(\i\w_n)= \int_0^{1/T} d\tau e^{\w_n\tau}
 \nonumber \\
& &\ \ \ \ \ \ \ \ \ \ \ \ \ \ \
 \times \langle T_\tau\left\{ c_{\k\s}^\dagger(\tau) c_{\k\s}(\tau) 
      c_{\k'\s'}^\dagger c_{\k'\s'} \right\} \rangle,
\end{eqnarray}
where $v_{\k\mu}^0(k)= \d \e_\k^0/\d k_\mu$ and
$\w_n= 2\pi T n$ is the even Matsubara frequency, and
${e}(>0)$ is the absolute value of the charge of an electron.
In the absence of the magnetic field, the analytic continuation 
from $K_{\k\s,\k'\s'}(\i\w_n)$ to $K_{\k\s,\k'\s'}(\w+\i\delta)$ 
has been performed by Eliashberg.
 \cite{Eliashberg}
According to him, the conductivity carried by the 
quasiparticles are given by
\begin{eqnarray}
 \sigma_{xx}= {e}^2 \sum_\k \left(-\frac{\d f}{\d\e} \right)_{\e_\k}
 z_\k v_{\k x} J_{\k x} \cdot \frac1{\Delta_\k},
 \label{eqn:sigma_xx}
\end{eqnarray}
where $f(\e)= 1/(1+{\rm e}^{(\e-\mu)/T})$.
In eq. (\ref{eqn:sigma_xx}) we have done the energy integration 
by assuming the relation $z_\k\Delta_\k\ll T$,
which is not always satisfied in HTSC as shown in \S VI, however.

\begin{figure}
\epsfxsize=70mm
\centerline{\epsffile{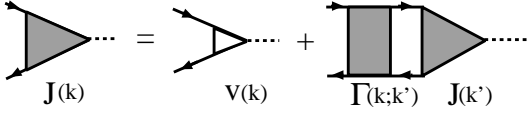}}
\caption{The BS equation for $J_{\k\mu}$.}
\label{fig:Bethe-Salpeter}
\end{figure}

In (\ref{eqn:sigma_xx}), $v_{\k x}$ and  $J_{\k x}$ are given by
\begin{eqnarray}
& & v_{\k x} = \frac{\d}{\d k_x}
 \left( \e_\k^0 + {\rm Re}\Sigma_\k(\w=0) \right), \\
& & J_{\k x} = v_{\k x} + \sum_{\k'} \int_{-\infty}^{\infty} 
 \frac{d\e}{4\pi\i}  {\cal T}_{\k\k'}(0,\e)
 \left| G_{\k'}(\e)\right|^2 \cdot J_{\k' x},
   \label{eqn:def_Jx}
\end{eqnarray}
where ${\cal T}_{\k\k'}(\e,\e')$ is the irreducible four point 
vertex, which plays an important role to treat the umklapp processes 
of conduction electrons.
 \cite{Yamada-Yosida} 
The total current ${\vec J}_{\k}$ is given by the solution of 
the Bethe-Salpeter (BS) equation
(\ref{eqn:def_Jx}), which is shown by Fig.
\ref{fig:Bethe-Salpeter}.
We note that ${\cal T}_{\k\k'}(\e,\e')$ 
is represented as ${\cal T}_{22}^{(0)}(\e,\e';\w=0)$
in the Eliashberg's paper.
 \cite{Eliashberg}

The Hall coefficient $R_{\rm H}$
under a weak magnetic field along $z$-axis $H$ is give by
\begin{eqnarray}
 R_{\rm H} = \frac{\s_{xy}/H}{\s_{xx}\s_{yy}}.
\end{eqnarray}
The analytic continuation for the normal Hall conductivity
$\s_{xy}$ due to the quasiparticle contribution in the presence of 
the magnetic field $H$ has been performed by 
Kohno and Yamada,
 \cite{Kohno-Yamada}
or Fukuyama {\it et al}.,
 \cite{Fukuyama}
in the gauge invariant manner.
According to them, 
in case there is the four-fold symmetry of the system, 
\begin{eqnarray}
\s_{\mu\nu}/H &=& -\e_{\mu\nu z}\cdot \frac12 {e}^3 
 \sum_\k \left(-\frac{\d f}{\d\e} \right)_{\e_\k}
  A_{\mu\nu}(\k) \frac{z_\k}{(\Delta_\k)^2} \nonumber \\
A_{\mu\nu}(\k) &=&
 v_{\k \mu} \left[ J_{\k \mu} \frac{\d J_{\k \nu}}{\d k_\nu} 
 -J_{\k \nu} \frac{\d J_{\k \mu}}{\d k_\nu} \right],
  \label{eqn:Kohno_Yamada}
\end{eqnarray}
where $\e_{xyz}=-\e_{yxz}=1$, 
reflecting the Onsagaer's reciprocity theorem.
Equation (\ref{eqn:Kohno_Yamada})
means $\s_{xy}/H\propto(\Delta_\k)^{-2}$,
whereas $\s_{xx}\propto(\Delta_\k)^{-1}$ by 
eq. (\ref{eqn:sigma_xx}).
Thus, $R_{\rm H}=$const. in the conventional Fermi liquid
at low temperatures.

The expression (\ref{eqn:Kohno_Yamada}) can be rewritten
into a simpler form, assuming that there is only
the symmetry with respect to the origin.
(see eq.(3.21) of ref. \cite{Kohno-Yamada}.)
\begin{eqnarray}
\s_{xy}/H &=& -\frac{{e}^3}{4} \sum_k \left(-\frac{\d f}{\d\e} \right)_{\e_\k}
 A_{\rm s}(\k) \frac{z_\k}{(\Delta_\k)^2} \nonumber \\
A_{\rm s}(\k) &=& A_{xy}(\k) + A_{yx}(\k) \nonumber \\
&=& \left[J_{\k x} \cdot \left( {\vec e}_z\times{\vec v}_\k \right)
    {\vec \nabla} \cdot J_{\k y}
   -\langle x \leftrightarrow y \rangle
 \right] 
 \nonumber \\
&=& |{\vec v}_\k| \cdot \left( {\vec J}_\k\times \frac{\d}{\d k_\parallel}
 {\vec J}_\k \right)_z 
 \nonumber \\
&=& |{\vec v}_\k| \cdot 
 |{\vec J}_\k|^2 \left( \frac{d\theta_J(\k)}{dk_\parallel} \right),
 \label{eqn:Kohno_Yamada2}
\end{eqnarray}
where $k_\parallel$ is the component of ${\vec k}$ along the vector
${\vec e}_\parallel(\k) = ({\vec e}_z\times{\vec v}_\k)/|{\vec v}_\k|$,
and is tangential to the FS at $\k$
because ${\vec v}_\k$ is perpendicular to the FS.
In eq. (\ref{eqn:Kohno_Yamada2}), $\theta_J(\k)$ is the angle between 
${\vec J}_\k$ and the $x$-axis, except for an arbitrary constant.
Contrary to $A_{\mu\nu}(\k)$, $A_{\rm s}(\k)$ introduced 
in eq. (\ref{eqn:Kohno_Yamada2}) is a scalar variable, i.e.,
independent of the choice of coordinates.
As a result, $\s_{xy}/H$ is also independent of the 
choice of coordinates, 
if only the reflection symmetry exists.
This property of $\s_{xy}/H$ has been proved so far only by
the Boltzmann transport theory.
 \cite{Tsuji,Zimann,Ong}

By using the relation $d \e_\k/dk_\mu = z_\k v_{\k\mu}$,
eq. (\ref{eqn:Kohno_Yamada2}) becomes 
\begin{eqnarray}
\sigma_{xy}/H =
 -\frac{{e}^3}{4} \oint_{\rm FS} dk_\parallel
 |{\vec J}_\k|^2 \left( \frac{d\theta_J(\k)}{dk_\parallel} \right) 
 \cdot \frac1{(\Delta_\k)^2}
  \label{eqn:s_xy_sign}
\end{eqnarray}
at sufficiently low temperatures.
In this line-integration, $\k$-point moves anti-clockwise 
along the FS.

Finally, we discuss on the Boltzmann transport theory:
The conductivity in the magnetic field ${\vec H}$ is given by
$\s_{\mu\nu} = e \sum_\k (-\d f/\d \e_\k) v_{\k\mu} \Phi_{\nu}(v_\k)$
for ${\vec E} \parallel {\hat e}_\nu$, where
$\Phi_{\nu}(v_\k)= (1- e\Delta_\k^{-1}
 ({\vec v}_\k \times {\vec H}) \cdot {\vec \nabla} )\cdot 
 ( e\Delta_\k^{-1} v_{\k\nu})$
up to the first order of ${\vec H}$ 
within the relaxation time approximation.
 \cite{Zimann}
As a result, the conductivity in this approximation,
$\sigma_{xx}^0$, is given by (\ref{eqn:sigma_xx}) by replacing
${\vec J}_\k$ with ${\vec v}_\k$.
In the same way, the Hall conductivity
within the relaxation time approximation is given by
\begin{eqnarray}
\sigma_{xy}^0/H = -\frac{{e}^3}{4} \oint_{\rm FS} dk_\parallel
 |{\vec v}_\k|^2 \left( \frac{d\theta_v(\k)}{dk_\parallel} \right) 
 \cdot \frac1{(\Delta_\k)^2},
  \label{eqn:s0_xy_sign}
\end{eqnarray}
where $\theta_v(\k)$ is the angle between 
${\vec v}_\k$ and the $x$-axis.
Thus, the sign of $\sigma_{xy}^0/H$ is determined by the sign of 
$d\theta_v(\k)/dk_\parallel$, which is nothing but
the curvature of the FS at $\k$.
 \cite{Tsuji,Zimann,Ong}

In the later sections, we calculate $J_{\k \mu}$
by solving the BS eq. (\ref{eqn:def_Jx}).
In the nearly AF Fermi liquid, we find that
${\vec J}_{\k}$ is no more perpendicular to the FS,
so $d\theta_J(\k)/dk_\parallel$ and $d\theta_v(\k)/dk_\parallel$ 
at the same $\k$ can be quite different, even in sign.
This is the reason why the Boltzmann approximation fails to 
reproduce the anomalous behavior of $R_{\rm H}$ in HTSC's.

\section{Vertex Corrections from ${\cal T}_{\k,\k'}(\e,\e')$}

\begin{figure}
\epsfxsize=60mm
\centerline{\epsffile{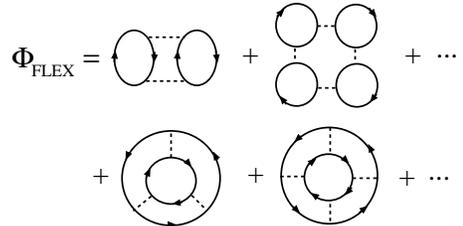}}
\caption{Each full line represents the dressed Green function
$G_\k(\w)$ and each broken line represents $U$. }
\label{fig:Luttinger}
\end{figure}

In this section, we study the vertex corrections for the current,
which is essential for the transport phenomena.
The self-energy in the FLEX theory, which is given by 
eq. (\ref{eqn:self}),
is also obtained by the functional derivative of 
$\Phi_{\rm FLEX}$ as
$ \Sigma_\k(\e)= \delta\Phi_{\rm FLEX}/\delta G_\k(\e)$,
where $\Phi_{\rm FLEX}$ is given by the closed skeleton diagrams
made of $G_\k(\e)$ and $U$, with a factor $1/n$ for $U^n$-diagrams.
The existence of $\Phi_{\rm FLEX}$, 
which is depicted in Fig. \ref{fig:Luttinger},
means that the FLEX theory is classified 
as a conserving approximation whose framework 
was constructed by Baym and Kadanoff,
 \cite{Baym-Kadanoff}
and Baym.
 \cite{Baym}
In the conserving approximation, the particle-hole transport
function $L$ is given as the solution of the BS 
equation, where the
irreducible particle-hole vertex 
$\Gamma_{\k\k'}(\e,\e')= \delta\Sigma_\k(\e)/\delta G_{\k'}(\e')$
is used as the kernel.
Then, the $L$ obtained in this way satisfies various
conservation laws automatically.
This is the reason why we call it the conserving approximation.

Significance of the conserving approximation
for the correlation functions 
is well recognized in various situations.
Conductivity is one typical quantity.
Within the conserving approximation, Yamada and Yosida show that 
the conductivity given by (\ref{eqn:sigma_xx})
diverges at finite temperatures 
in the absence of the umklapp processes,
reflecting the momentum conservation law.
Their work also shows that the vertex correction from ${\cal T}_{\k\k'}(\e,\e')$ 
in (\ref{eqn:def_Jx}), which is neglected within the Boltzmann theory,
is indispensable to treat the umklapp scattering 
processes of electrons self-consistently.

\begin{figure}
\epsfxsize=70mm
\centerline{\epsffile{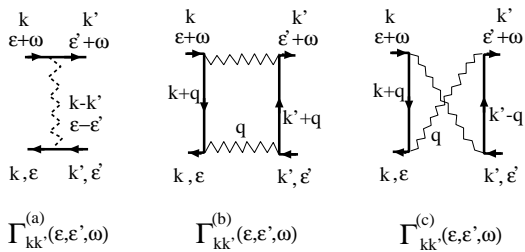}}
\caption{The irreducible four-point vertexes,
 which are sufficient for the conserving approximation.
 In (a), the wavy line represents $V(k-k')$.
 In (b) and (c), the two wavy lines represent $W(q)$
 and the $q$-summation should be taken.} 
\label{fig:vertex}
\end{figure}

The irreducible particle-hole vertex $\Gamma_{\k\k'}(\e,\e')$
within the FLEX theory is shown in Fig. \ref{fig:vertex}.
\begin{eqnarray}
\Gamma_{\k\k'}(\e_n,\e_{n'};\w_l) = \Gamma^{(a)}
 +\Gamma^{(b)}+\Gamma^{(c)},
\end{eqnarray}
where $\e_n= (2n+1)\pi T$ is the odd Matsubara frequency,
and $\w_l= 2l\pi T$ is the external frequency.
We put the external momentum zero.
$\Gamma^{(a)}$-$\Gamma^{(c)}$ are given by
\begin{eqnarray}
& &\Gamma_{\k\k'}^{(a)}(\e_n,\e_{n'};\w_l) =
 V_{\k-\k'}(\e_n-\e_{n'}),
 \label{eqn:A1} \\
& &\Gamma_{\k\k'}^{(b)}(\e_n,\e_{n'};\w_l) = -T\sum_{\q,l'} 
  W_\q(\w_{l'},\w_{l'}-\w_l)  \nonumber \\
& &\ \ \ \ \ \ \ \ 
 \times G_{\k+\q}(\e_n+\w_{l'}) G_{\k'+\q}(\e_{n'}+\w_{l'}) , 
 \label{eqn:A2} \\
& &\Gamma_{\k\k'}^{(c)}(\e_n,\e_{n'};\w_l) = -T\sum_{\q,l'} 
  W_\q(\w_{l'},\w_{l'}+\w_l) \nonumber \\
& &\ \ \ \ \ \ \ \ 
 \times G_{\k+\q}(\e_n+\w_{l'}+\w_l) G_{\k'-\q}(\e_{n'}-\w_{l'}) ,
 \label{eqn:A3} 
\end{eqnarray}
where $V_\k(\e_n)$ is given by (\ref{eqn:def_V}),
and we have introduced $W_\k(\w_l,\w_{l'})$ as
\begin{eqnarray}
W_\k(\w_l,\w_{l'})
   &=& \frac32 U^2 \left( U\chi_\k^{s}(\w_l)+1 \right)
                   \left( U\chi_\k^{s}(\w_{l'})+1 \right)
    \nonumber \\
   &+& \frac12 U^2 \left( U\chi_\k^{c}(\w_l)-1 \right)
                    \left( U\chi_\k^{c}(\w_{l'})-1 \right)
    \nonumber \\
   & &- U^2.
 \label{eqn:W-def}
\end{eqnarray}
These three irreducible vertices
$\Gamma^{(a)}$, $\Gamma^{(b)}$ and $\Gamma^{(c)}$
are sufficient for the conserving approximation.
In literatures, (a) process is called Maki-Thompson (MT) term, and
(b) and (c) are called Aslamazov-Larkin (AL) terms, respectively.

In order to solve the BS equation (\ref{eqn:def_Jx}),
we have to obtain the functional form of the irreducible vertex 
${\cal T}_{\k\k'}(\e,\e')$ in (\ref{eqn:def_Jx}).
For this purpose, we perform the analytic continuation of
$\Gamma^{(a)}$, $\Gamma^{(b)}$ and $\Gamma^{(c)}$ 
with respect to $\e_n$ and $\e_{n'}$ in Appendix A.
Thus, we obtain the vertex corrections for the current
by replacing ${\cal T}_{\k\k'}(0,\e)$ in eq. (\ref{eqn:def_Jx})
with ${\cal T}^{(a{\mbox{-}}c)}_{\k\k'}(0,\e)$ given by eqs.
(\ref{eqn:Ta})-(\ref{eqn:Tc}).

At first, the contribution coming from ${\cal T}^{(a)}_{\k\k'}$
is given by
\begin{eqnarray}
{\mit\Delta} J_{\k\mu}^{a} &=& \sum_{\k'} \int \frac{d\e'}{2}
 \left( {\rm cth}\frac{\e'}{2T} - {\rm th}\frac{\e'}{2T} \right)
 {\rm Im}V_{\k'-\k}(\e'+\i\delta) 
 \nonumber \\
& & \times \rho_{\k'}(\e') \frac{1}{\Delta_{\k'}(\e')} J_{\k'\mu},
   \label{eqn:DJa}
\end{eqnarray}
where we put $\e=0$, and
we have used the relation
$|G_\k(\e)|^2= \pi\rho_\k(\e)/\Delta_{\k}(\e)$.
This vertex corrections play an important role in the
singular behavior of the Hall coefficient in HTSC's,
which will be discussed in \S V.

Next, we consider the correction terms
come from ${\cal T}^{(b)}_{\k\k'}$ and ${\cal T}^{(c)}_{\k\k'}$.
Approximately, they are given by
\begin{eqnarray}
{\mit\Delta}J_{\k\mu}^{b,c} &=&  
 \frac{\pi}{4} \sum_{\k'}\int d\e
 \left( {\rm cth}\frac{\e}{2T}-{\rm th}\frac{\e}{2T} \right)
 \sum_\q \int d\w W_\q(\w)
  \nonumber \\
& &\times 
 \left( {\rm th}\frac{\w+\e}{2T}  - {\rm th}\frac{\w}{2T} \right)
 \rho_{\k+\q}(0) \rho_{\k'+\q}(0) 
  \nonumber \\
& &\times 
  \frac{\rho_{\k'}(0)}{\Delta_{\k'}(0)} J_{\k'\mu} ,
  \label{eqn:DJbc} 
\end{eqnarray}
where $W_\q(\w)$ is introduced by eq. (\ref{eqn:W-def2}).
Equation (\ref{eqn:DJbc}) is derived by neglecting
the $\w$-dependences of $\rho_\k(\w)$ and $\Delta_\k(\w)$,
which will be allowed qualitatively because 
only the regions $|\w|,|\e|\simle \min\{ T,\w_{\rm sf} \}$ 
are important in the $\w,\e$-integrations in eq. (\ref{eqn:DJbc}).
The variable change $\k' \rightarrow -\k'$ is performed
for ${\mit\Delta}{\vec J}_{\k}^{c}$.
Strictly speaking, however,
${\mit\Delta}{\vec J}_{\k}^{b}(\w)$ and 
${\mit\Delta}{\vec J}_{\k}^{c}(\w)$ are not equal.

Now, we show that ${\mit\Delta}J_{\k\mu}^{b,c}$ 
is negligible in the case of $\xi^2\gg1$ and $\Q=(\pi,\pi)$.
In this case, the leading contributions in the $\q$-summation 
in eq. (\ref{eqn:DJbc}) come only from $\q\sim \Q$.
In the $\k'$-summation of eq. (\ref{eqn:DJbc}), 
there is a cancellation between the contributions from
$\k'$ and $-\k'$ if we put $\q=\Q$, because $\k'+\Q$ and $\k'-\Q$ 
are the same in the momentum space.
As a result, ${\mit\Delta}J_{\k\mu}^{b,c}$ is expected to be
negligible compared with ${\mit\Delta}J_{\k\mu}^{a}$.
This statement becomes rigorous in the case of $\xi^2\rightarrow\infty$.
In Appendix B, we show this cancellation in the two AL terms
explicitly by the numerical calculations.
This is one of the main results of this paper.

\section{Analysis for the Bethe-Salpeter (BS) Equation for $J_{\k \mu}$}

The aim of this section is to give the qualitative understanding
of the mechanism for the temperature dependence of $R_{\rm H}$ in HTSC's.
We try an analytical approach
to solve the BS equation (\ref{eqn:def_Jx})
for $J_{\k \mu}$, by neglecting the AL-terms.
To simplify the discussion,
we assume that $\Q=(\pi,\pi)$ and 
the MBZ-boundary lies on 
$(\pm\pi,0)$ and $(0,\pm\pi)$,
which is realized in YBCO experimentally.

For a qualitative discussion, 
we use the phenomenological expression for $\chi_\q^s(\w)$,
eq. (\ref{eqn:kai_qw}), and neglect other terms
in the definition of $V_\q(\w)$ in eq. (\ref{eqn:def_V}).
We introduce the function
$H(x)= 1/x - 2\psi(x+1) + 2\psi(x+1/2)$,
where $\psi(x)$ is the di-Gamma function.
Then, the imaginary part of the self-energy is given by
\begin{eqnarray}
\Delta_\k&=& \sum_\q\int \frac{d\e}{2}
 \left[ {\rm cth}\frac{\e}{2T}-{\rm th}\frac{\e}{2T} \right] \cdot
 {\rm Im} V_\q(\e+\i\delta) \cdot \rho_{\k-\q}(\e)
  \nonumber \\
&=& \frac{3U^2}{4} \sum_\q \chi_Q \w_{\rm sf}
 \cdot H\left(\frac{\w_\q}{2\pi T} \right) \cdot \rho_{\k-\q}(0) 
 \label{eqn:Delta} \\
& &H\left(\frac{\w_\q}{2\pi T} \right)
 \approx  \frac{(\pi T)^2}{\w_\q(\w_\q + \pi T/2)} 
 \label{eqn:H-approx} 
\end{eqnarray}
where 
$\w_\q= \w_{\rm sf} + \w_{\rm sf}\xi^2(\q-\Q)^2$.
 \cite{Hall-Pines}
We see that $H({\w_\q}/{2\pi T})$ takes a large value only when
$|\Q-\q|\simle \xi^{-1}$, and it is negligibly small elsewhere.
The approximate form of eq. (\ref{eqn:Delta}) 
with eq. (\ref{eqn:H-approx}) is also obtained
by extrapolating the results of both limits, 
$\w_{\rm sf} \gg T$ and $\w_{\rm sf} \ll T$,
where the $\e$-integration in eq (\ref{eqn:Delta}) becomes easier.

\begin{figure}
\epsfxsize=70mm
\centerline{\epsffile{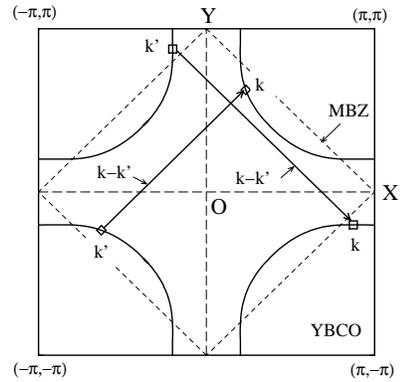}}
\caption{The relation between $\k$ and $\k'$.
 Both points locate on the FS.}
\label{fig:FS-delk}
\end{figure}

Next, we examine the vertex correction for the current, eq. (\ref{eqn:DJa}).
We stress that Im$V_\k(\w)$ appearing in eqs.
(\ref{eqn:DJa}) and (\ref{eqn:Delta}) are same, 
which is ensured in the conserving approximation.
We can show that
\begin{eqnarray}
{\mit\Delta}J_{\k \mu}
&=& \frac{3U^2}{4} \sum_\q
 \chi_Q \w_{\rm sf} \cdot H\left(\frac{\w_\q}{2\pi T} \right) 
 \cdot \frac{\rho_{\k-\q}(0)}{\Delta_{\k-\q}}\cdot J_{\k-\q \mu} 
 \label{eqn:JJJ}
\end{eqnarray}
Comparing eq. (\ref{eqn:JJJ}) with eq. (\ref{eqn:Delta}), 
and noticing that $H({\w_\q}/{2\pi T})$
is negligibly small for $|\Q-\q|\simge 1/\xi$,
we get
\begin{eqnarray}
{\mit\Delta}{\vec J}_{\k} 
 &\approx& \ \langle \ {\vec J}_\q \ \rangle_{|\q-\k'|<1/\xi}
   \nonumber \\
 &\approx& \ {\vec J}_{\k'} \cdot 
  \langle \ \cos(\theta_J(\q)\!-\!\theta_J(\k')) \ \rangle_{|\q-\k'|<1/\xi} ,
 \label{eqn:Jdash}
\end{eqnarray}
where $\k$, $\k'$ and $\q$ are on the FS.
Here, we have introduced $\k'$ on the FS so that 
$(k_x',k_y')=(k_y,k_x) \cdot {\rm sign}( -k_x\!\cdot\! k_y)$,
as shown in Fig. \ref{fig:FS-delk}.
We see the relation $\k-\k'\approx\Q$ is satisfied
in the momentum space.
Here we assume $|\Q-(\k-\k')|\simle \xi^{-1}$ around the cold spot
because it seems to be satisfied in the present numerical calculation
by the FLEX theory.
Thus, we obtain a simplified BS equation,
\begin{eqnarray}
{\vec J}_\k= {\vec v}_\k + \a_\k\cdot{\vec J}_{\k'},
 \label{eqn:BS_simple}
\end{eqnarray}
where $\a_\k \approx (1-c/\xi^2)<1$ and $c\sim O(1)$ is a constant.
$\a_\k$ takes the maximum value around hot spots.

\begin{figure}
\epsfxsize=60mm
\centerline{\epsffile{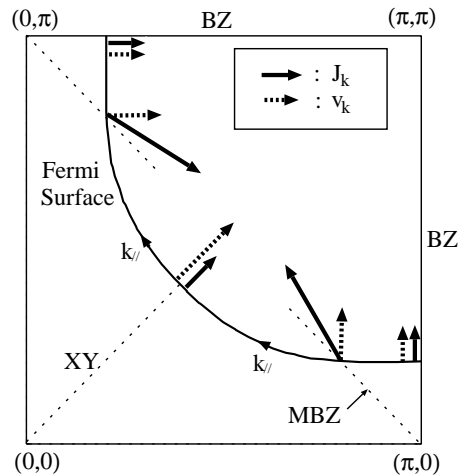}}
\caption{Schematic behaviors of ${\vec J}_\k$ and ${\vec v}_\k$.
 Contrary to ${\vec v}_\k$, ${\vec J}_\k$ is not 
 perpendicular to the FS.
 For example, $(d\theta_J/dk_{\parallel})<0$ on the XY-axis, 
 and $(d\theta_J/dk_{\parallel})>0$ on the BZ-boundary.}
 (see eq. (\ref{eqn:Jy}).)
\label{fig:J-schematic}
\end{figure}

Then, eq. (\ref{eqn:BS_simple}) is easily solved as
\begin{eqnarray}
{\vec J}_{\k}= \frac 1{1-\a_\k^2} 
 \left( {\vec v}_{\k} + \a_\k \cdot {\vec v}_{\k'} \right),
  \label{eqn:Jy}
\end{eqnarray}
Equation (\ref{eqn:Jy}) means that 
${\vec J}_\k$ is not parallel to ${\vec v}_\k$.
For example,
(i) at $k_x=k_y$, 
${\vec J}_\k = {\vec v}_\k/(1+\a_\k) \sim \frac12 {\vec v}_\k$
is satisfied.
(ii) Near the MBZ-boundary,
${\vec J}_\k \approx (\xi^2/2c)( {\vec v}_{\k} + {\vec v}_{\k'} )$,
which is nearly parallel or perpendicular to ${\Q}$.
(iii) On the BZ-boundary, 
${\vec J}_\k \parallel {\vec v}_\k$ and
${\vec J}_\k \approx {\vec v}_\k$ is realized because
the contribution from $\k'$-point
cancels out with that from $-\k'$-point approximately,
due to the fact that $|\k-\k'|=|\k+\k'|\approx|\Q|$ in the momentum space.
Thus, we should put $\a_\k=0$ in eq. (\ref{eqn:Jy})
on the BZ-boundary.
These behaviors of ${\vec J}_\k$ together with ${\vec v}_\k$
are shown schematically in Fig. \ref{fig:J-schematic},
which is confirmed by the numerical calculation in \S IV B.
Physically, this peculiar behavior of ${\vec J}_\k$ 
comes from the multiple backward scattering between 
$\k$ and $\k'$ caused by the AF fluctuations.

Apparently, 
${\vec J}_{\k}$ shows the divergent behavior
near the AF phase boundary where $\a_\k\approx1$ is realized.
Now we stress the importance of the conservation approximation
to get the correct $\a_\k$.
For instance, 
we always get $\a_\k = \infty$ 
if we replace $|G_\k(\e)|^2$ with $|G_\k^0(\e)|^2$ in 
eq. (\ref{eqn:def_Jx}), which leads to the divergence of ${\vec J}_\k$.

First, we consider the conductivity $\s_{xx}$.
According to (\ref{eqn:sigma_xx}), 
$\s_{xx}$ is given by the averaged value of $v_{\k x}\cdot J_{\k x}$
over the FS, that is,
\begin{eqnarray}
& &v_{\k x}\cdot J_{\k x} = \frac 1{1-\a_{\k}^2}
 \left\{ |v_{\k x}|^2 - \a_{\k} |v_{\k x} v_{\k y}| \right\} .
 \label{eqn:vJ}
\end{eqnarray}
We see that
eq. (\ref{eqn:vJ})$\approx  |v_{\k x}|^2/(1+\a_{\k})$ 
is realized around the cold spots in YBCO and LSCO,
where $|v_{\k x}|\approx |v_{\k y}|$ is satisfied.
This means that the conductivity $\s_{xx}$ 
is smaller than that given by the Boltzmann approximation,
due to the vertex corrections for the current given
by ${\cal T}_{\k\k'}(\e,\e')$.
This is confirmed by the numerical calculations in \S VI.

Next, we discuss the $\s_{xy}$.
By using eq. (\ref{eqn:Jy}), $dJ_{\k x}/dk_\mu$ is given by
\begin{eqnarray}
\frac{d}{d k_\mu} J_{\k x}
&=& \frac{\mp \beta_{\k,\mu}}{1-\a_\k^2} \cdot v_{\k y}
 +\frac{2\a_\k \beta_{\k,\mu}}{1-\a_\k^2} \cdot J_{\k x}
 \nonumber \\
 & &+ \frac 1{1-\a_\k^2} \cdot \left( \frac{dv_{\k x}}{dk_\mu} 
    \mp \a_\k \frac{dv_{\k y}}{dk_\mu} \right) ,
\end{eqnarray}
where $\beta_{\k,\mu}\equiv d\a_\k/dk_\mu$.
Hereafter, the $\mp$ in equations is equal to 
${\rm sign}(-k_x\!\cdot\!k_y)$.
By using eq. (\ref{eqn:JJJ}), which gives the definition of $\a_\k$,
then $\beta_{\k,\mu}$ is given as
\begin{eqnarray}
\beta_{\k,\mu} 
\approx -\xi^2 (\Q-(\k-\k'))_\mu
 \label{eqn:beta_def}
\end{eqnarray}
when $|\Q-(\k-\k')|\simle 1/\xi$ is satisfied.
We see that 
$\beta_{\k,\mu}$ is positive when ${\vec k}+ \e\cdot{\vec e}_\mu$ 
is closer to the MBZ-boundary than ${\vec k}$ is 
(${\vec e}_\mu$ is a unit vector along $\mu$-direction,
 and $\e>0$ is a small constant.), and vice versa.

As a result, $A_{\rm s}(\k)$ introduced in 
eq. (\ref{eqn:Kohno_Yamada2}) is given by
\begin{eqnarray}
\frac{1}{|{\vec v}_\k|} A_{\rm s}(\k) 
&=& \frac 1{1-\a_\k^2} \left({\vec v}_{\k} \times 
  \frac{d}{d k_\parallel} {\vec v}_{\k} \right)_z
  \nonumber \\
& &+ \frac {\mp\beta_{\k,\parallel}}{(1-\a_\k^2)^2} 
 \left[ v_{\k x}^2 - v_{\k y}^2 \right],
 \label{eqn:vJJ}
\end{eqnarray}
where the momentum $k_\parallel$ is tangential 
to the FS, and is along the anti-clockwise direction.
Thus, $\s_{xy}/H$ is enhanced by the factor,
$1/(1-\a_\k^2) \propto \xi^2$ or
$\beta_{\k,\parallel}\propto \xi^2$,
contrary to the case of $\s_{xx}$.
The first term of (\ref{eqn:vJJ}) is proportional to
the contribution given by the Boltzmann transport theory,
whose sign is determined by the curvature of the FS.
It takes a larger value inside of the MBZ. (see fig. \ref{fig:FS-U}.)
On the other hand, the second term of (\ref{eqn:vJJ})
is negative inside of the MBZ, and is positive outside of it.
This term is dominant outside of the MBZ
because $|\beta_{\k,\parallel}|\gg1$ and 
$|v_{\k x}^2-v_{\k y}^2|\sim |{\vec v}_\k|^2$ is satisfied there.
We notice that $\beta_{\k, \parallel}=0$ on the MBZ-boundary and 
on the XY-axis.

The obtained results in HTSC's with the strong AF fluctuations
are summarized as follows (see Fig. \ref{fig:J-schematic}.) :
\begin{itemize}
\item The portion of the FS inside 
of the MBZ gives rise a positive contribution to $R_{\rm H}$.
In other wards, 
$A_{\rm s}(\k) \propto (d\theta_J(\k)/dk_\parallel)<0$ 
inside of the MBZ.
\item The outside part of the MBZ 
gives rise a positive contribution to $R_{\rm H}$.
In other wards, 
$(d\theta_J(\k)/dk_\parallel)>0$ there.
\end{itemize}
In the above $\theta_J(\k)$ is introduced in eq. (\ref{eqn:s_xy_sign}).
This change of the sign of $R_{\rm H}$ never occurs within the
Boltzmann approximation because 
$(d\theta_v(\k)/dk_\parallel)<0$ everywhere.

The Hall coefficient will be determined by the region near
the cold spots because of the factor $(\Delta_\k)^{-2}$ 
in eq. (\ref{eqn:s_xy_sign}).
As shown in Fig. \ref{fig:FS-hotcold}, 
the cold spots locate inside (outside) of the MBZ
in the case of YBCO (NCCO).
Thus, we understand the reason why 
$R_{\rm H}>0$ in the hole doped systems,
and why the sign of $R_{\rm H}$ changes in
the electron-doped systems.

In conclusion,
the $\s_{xy}/H$ is proportional to $\xi^2\propto 1/T$
both in the hole-doped systems and 
in the electron-doped systems.
We find that the $T$-dependence of $\s_{xx}$ and 
$\s_{xy}/H$ for a system with the strong AF fluctuations are
\begin{eqnarray}
& &\s_{xx}   \sim \xi^0/\Delta_\k,
 \nonumber \\
& &\s_{xy}/H \sim \pm\xi^2/\Delta_\k^2,
 \label{eqn:RH-est} \\
& &R_{\rm H} \sim \pm\xi^2 \sim \pm\chi_Q,
 \nonumber 
\end{eqnarray}
where $+(-)$ is for hole-doping (electron-doping) case.
The factor $\xi^2$ comes from the vertex corrections 
for the current introduced in this paper, which does not
appear within the Boltzmann approximation.
We will confirm this analysis by numerical calculations
based on the FLEX theory in the next section.

\section{Numerical Results}

\subsection{One-Particle Properties and Magnetic Properties 
 obtained by the FLEX approximation}

Here, we show electronic properties 
obtained by the FLEX approximation.
In this section,
we use $U=8|t_0|$ for YBCO in numerical calculations, 
considering that the band width $W$ is $8|t_0|$.
On the other hand, we use
$U=6|t_0|$ for LSCO, and $U=5.5|t_0|$ for NCCO,
to reduce the Stoner factor in the FLEX calculation,
$\a_{\rm S}= \max_\q\{U\cdot\chi_\q(0)\}$.
We have checked the numerical results do not depend on
$U$ qualitatively.

In this section, we put $|t_0|=1$.
Then, $T=0.1$ will corresponds to $\sim 600$K
because $|t_0|\sim0.5$eV in the LDA band calculation.
In the present calculation, 4096 $\k$-point meshes 
and 256 Matsubara frequencies are used.
By solving the linearized Eliashberg equations,
we obtain $T_{\rm c}\approx 0.02$ ($\sim 120$K) for YBCO and LSCO at $n=0.85$,
which is close to the $T_{\rm c}$'s reported 
by the previous studies based on the FLEX approximation.
 \cite{Monthoux-Scalapino,Dahm-Tewordt,Takimoto,d-p-model}
Also, $T_{\rm c}\approx 0.01$ for NCCO at $n=1.15$.
We find that the symmetry of the superconducting state is 
$d_{x^2-y^2}$-like in all cases.

\begin{figure}
\epsfxsize=65mm
\centerline{\epsffile{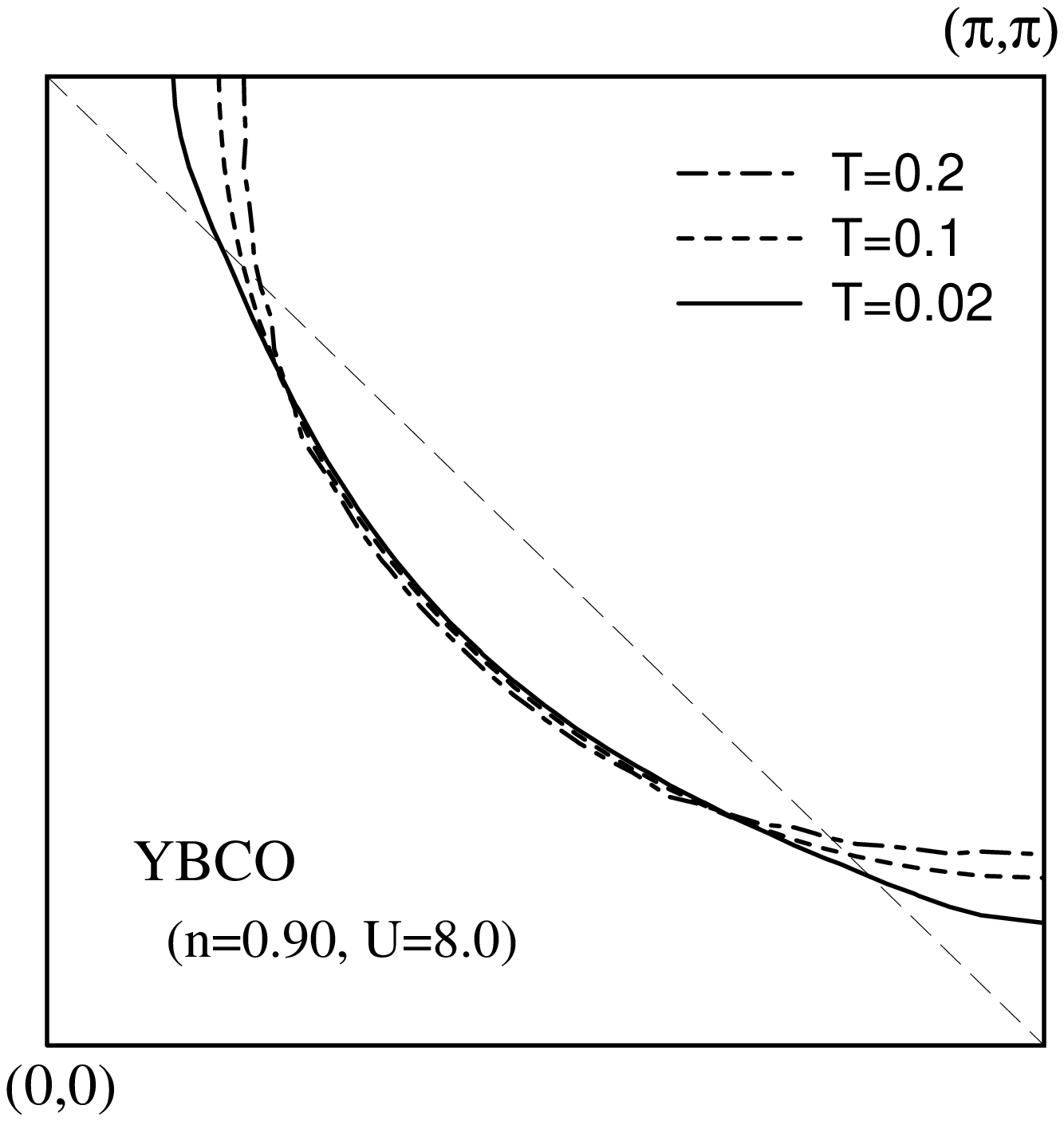}}
\epsfxsize=65mm
\centerline{\epsffile{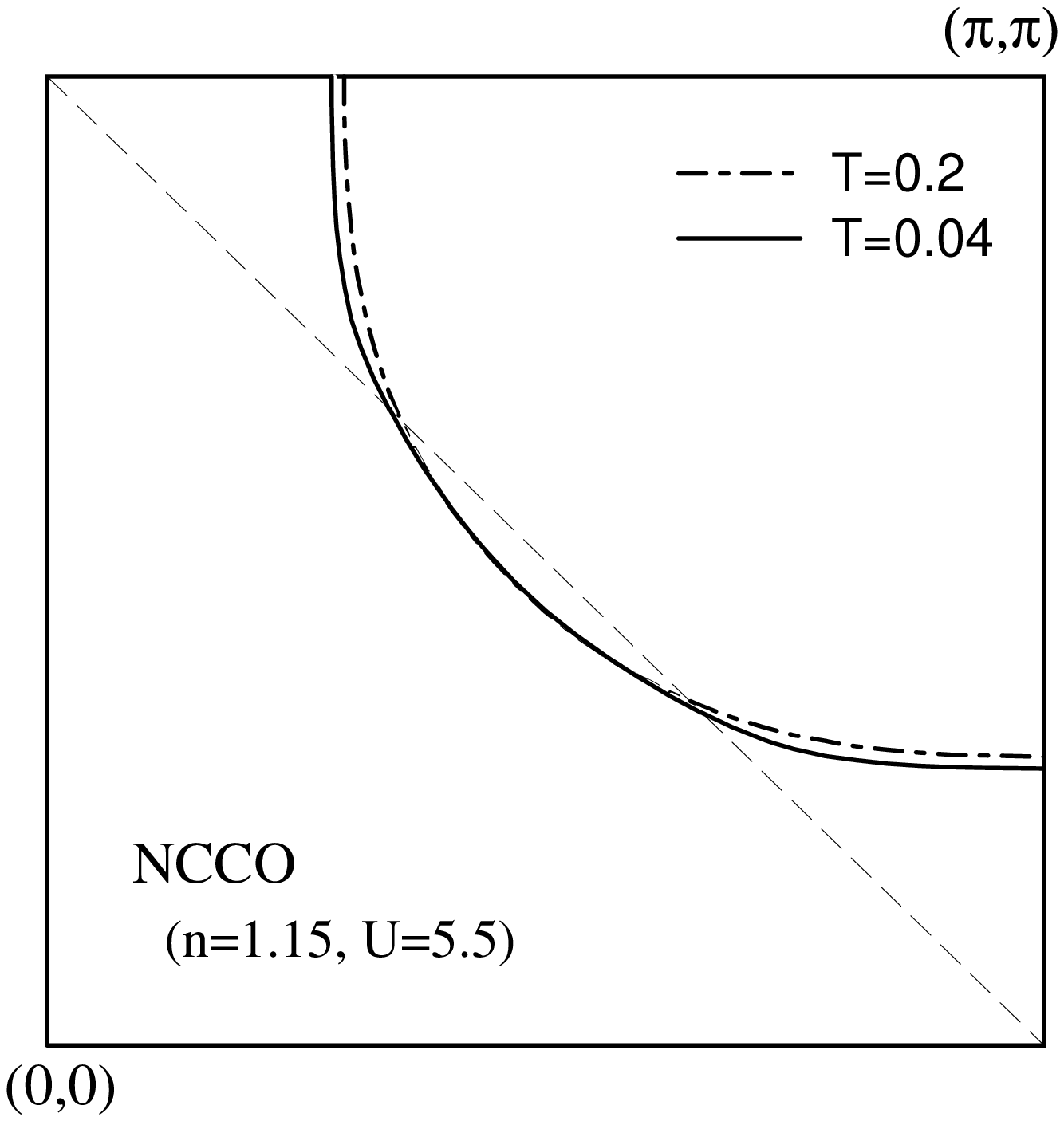}}
\epsfxsize=65mm
\centerline{\epsffile{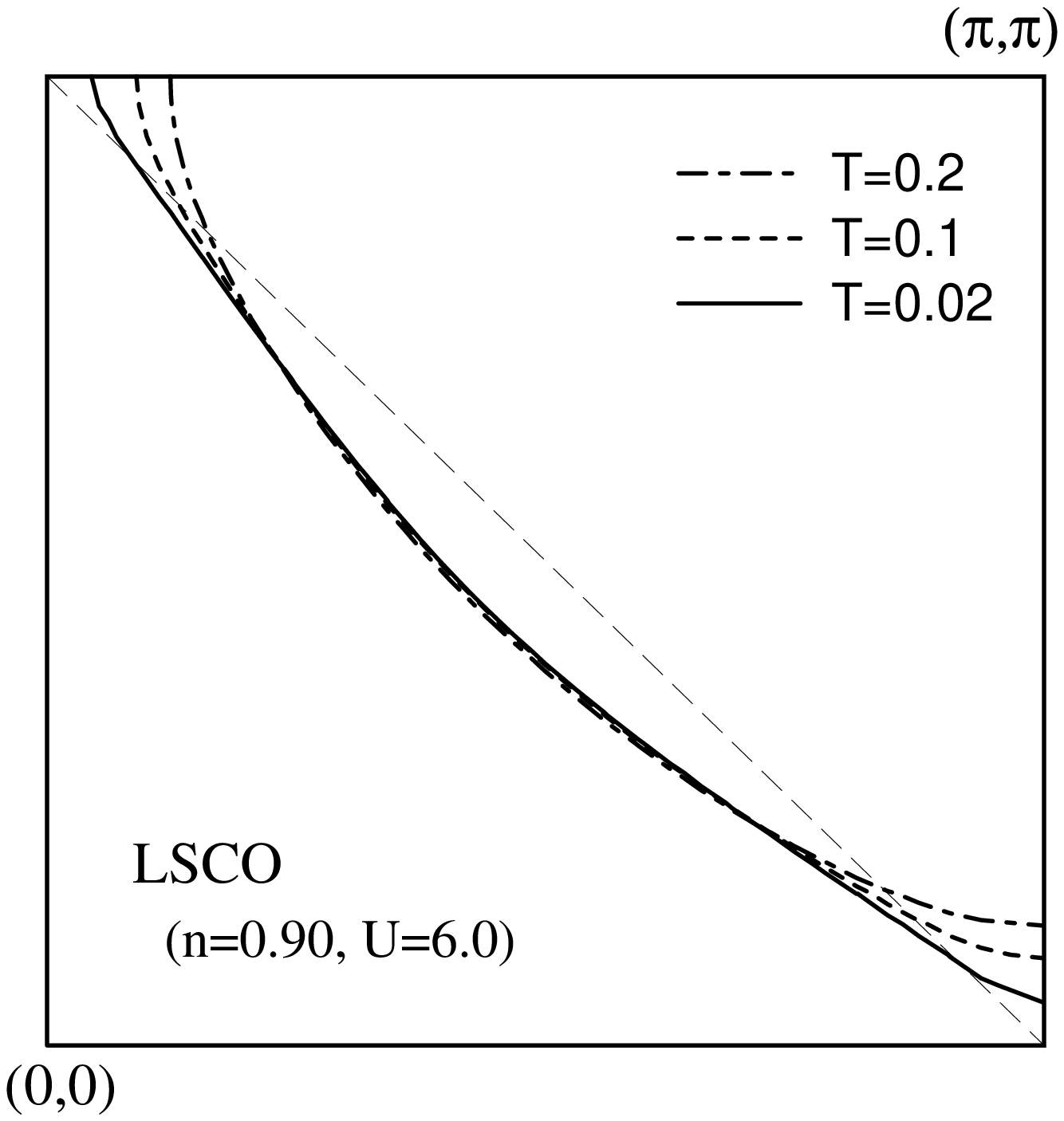}}
\caption{The temperature dependence of the interacting FS's
 for various compounds.
 It is small and negligible in the case of $U=0$.}
\label{fig:FS-T}
\end{figure}

Figure \ref{fig:FS-T} shows the temperature dependence 
of the FS's for YBCO, NCCO and LSCO, respectively.
They are determined by the relation
$\e_\k^0 + {\rm Re}\Sigma_\k(0)=\mu$.
At low temperatures,
the curvature of the interacting FS becomes smaller,
which is more prominent in YBCO and LSCO.
The deformation of the FS originates from Re$\Sigma_\k(0)$,
whose analytical expression at zero temperature
is given by ref.
 \cite{Yanase}
within the spin-fluctuation model.
According to it, the sign of Re$\Sigma_\k(0)$ is equal to 
${\rm sign}(\mu-\e_{\k-\Q})$ approximately, 
which moves the FS towards the MBZ-boundary.
As the temperature increases, the FS becomes closer
to the non-interacting one because $\xi$ decreases.
This temperature dependence of the FS
should make $|R_{\rm H}|$ smaller as the temperature decreases,
because its curvature around the cold spots decreases then.
We examine this effect numerically in the next subsection.

Moreover, in the FLEX approximation, the
flat-band structure ({\it i.e.}, extended saddle point)
is created around the van-Hove singularity points, 
$(\pm\pi,0)$ and $(0,\pm\pi)$,
because of the renormalization effect by $1/z_\k \simle 10$.
 \cite{d-p-model}
This is also the origin of the  
sensitive temperature dependence of the FS in YBCO and LSCO
shown in Fig. \ref{fig:FS-T}.
This flat-band structure is actually observed by ARPES experiments.
 \cite{flat-band}

Next, we consider the $\q$-dependence of the 
static magnetic susceptibility $\chi_\q^s(0)$,
given by eq. (\ref{eqn:chi_s}).
Because it does not contain the vertex corrections
required in the conservation approximation,
it gives a slightly over-estimated value
in the under-doped region.
 \cite{Dahm-Tewordt}
In reality, the observed $\chi_\q^s(0)$ by the neutron 
diffraction experiments can not be expressed by 
the simple spin-fermion model in eq.(\ref{eqn:kai_qw}):
For YBCO, $\chi_\q^s(0)$ shows a peak 
around $\q \approx (\pi,\pi)$.
 \cite{YBCO-neutron}
On the other hand, it is incommensurate 
for La$_{2-\delta}$Sr$_\delta$CuO$_4$, 
and shows a peak around
$\q \approx ((1-\delta)\pi,\pi), (\pi,(1-\delta)\pi)$
for $0.18\simge\delta\simge0.05$.
 \cite{LSCO-neutron}

\begin{figure}
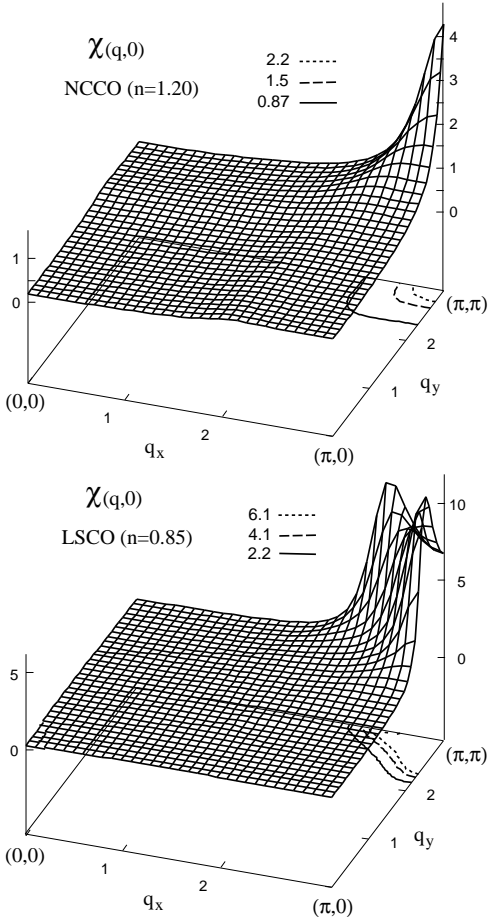

\epsfxsize=65mm
\centerline{\epsffile{kai.010-YBCO64-U8-n085.eps}}
\epsfxsize=65mm
\centerline{\epsffile{kai.008-Nd64-U55-n120.eps}}
\epsfxsize=65mm
\centerline{\epsffile{kai.010-LSCO-S64-U6-n085.eps}}
\caption{The ($q_x$,$q_y$)-dependence of $\chi_\q^s(0)$
 obtained by the FLEX approximation.
}
\label{fig:kai}
\end{figure}

Figure \ref{fig:kai} shows the calculated $(q_x,q_y)$-dependence of 
$\chi_\q(0)$
for YBCO ($n=0.90$, $T=0.02$), NCCO ($n=1.20$, $T=0.02$)
and LSCO ($n=0.85$, $T=0.06$), respectively.
We see that $\chi_\q(0)$ is commensurate
for YBCO and NCCO, which is also consistent with 
neutron diffraction experiments.
In the case of LSCO, $\chi_\q(0)$ shows an incommensurate 
structure at low temperatures.
At $n=0.85$, the peaks locate at
$\q= (0.83\pi,\pi),(\pi,0.83\pi) $ at $T=0.02$, which is consistent
with experiments, and
it becomes commensurate for $T\ge0.08$.
In conclusion, main characters of $\chi_\q^s(0)$ 
for each compounds are reproduced well by the FLEX calculation
with appropriate set of parameters, ($t_0\sim t_2$,$U$).

\begin{figure}
\epsfxsize=65mm
\centerline{\epsffile{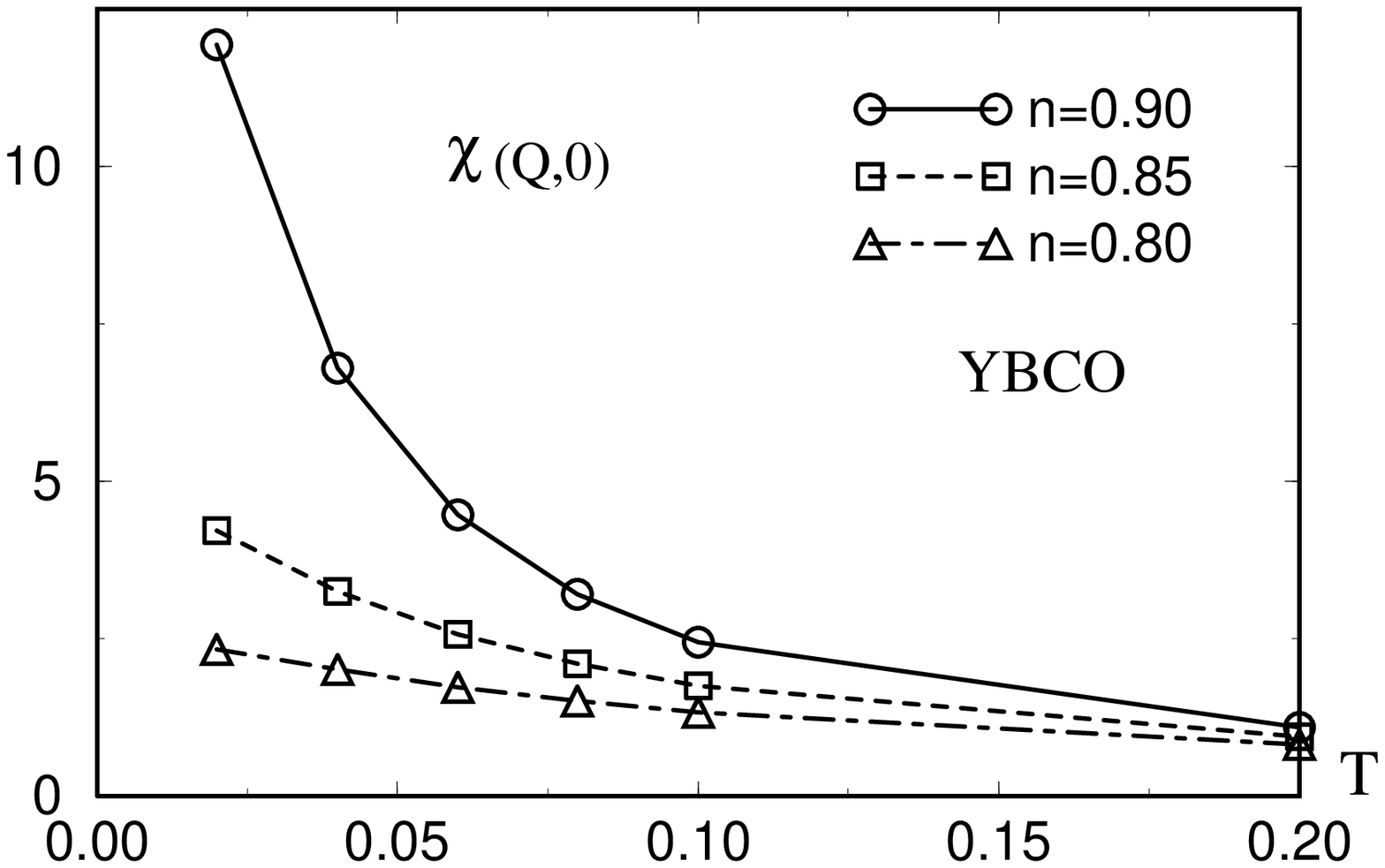}}
\epsfxsize=65mm
\centerline{\epsffile{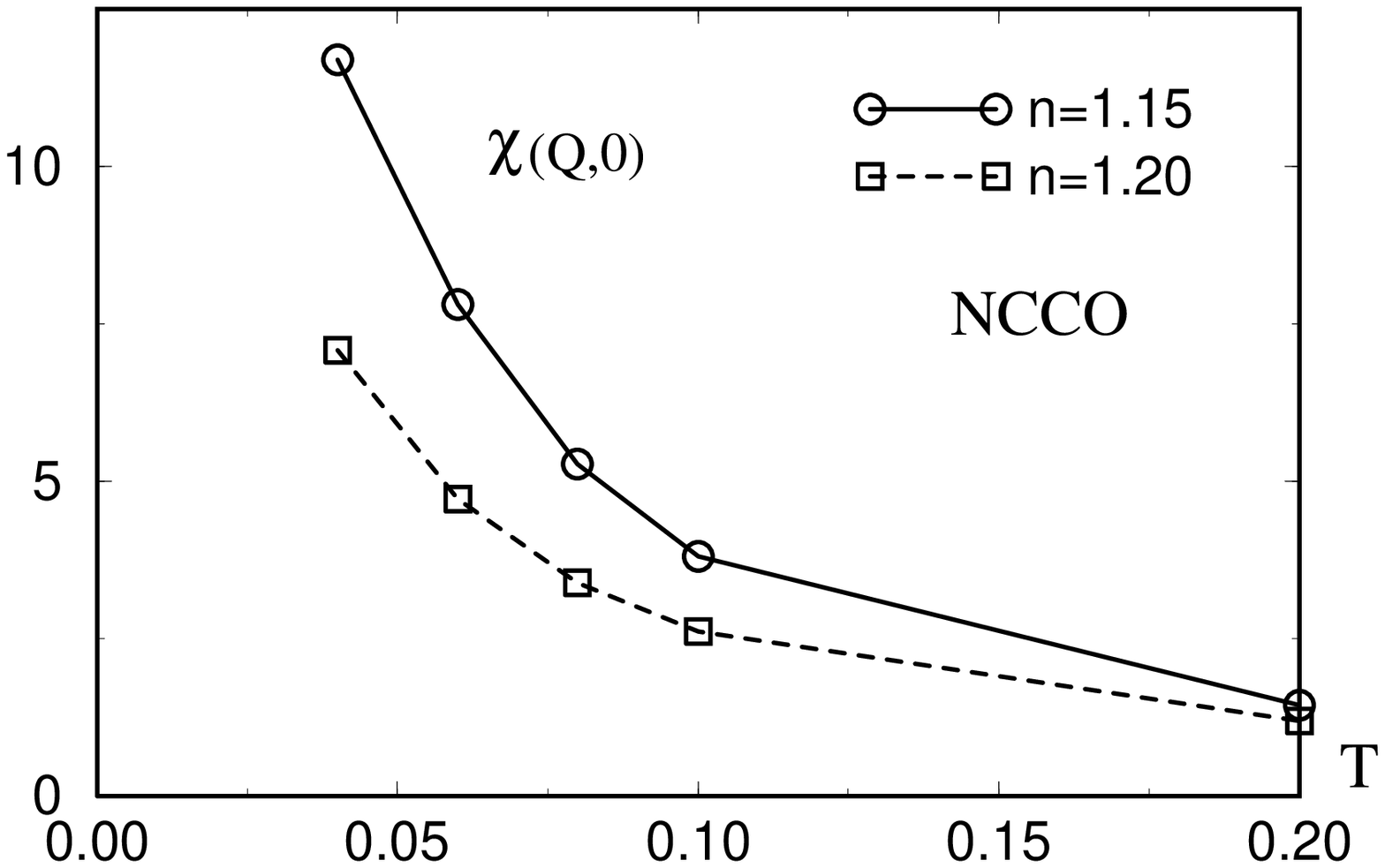}}
\epsfxsize=65mm
\centerline{\epsffile{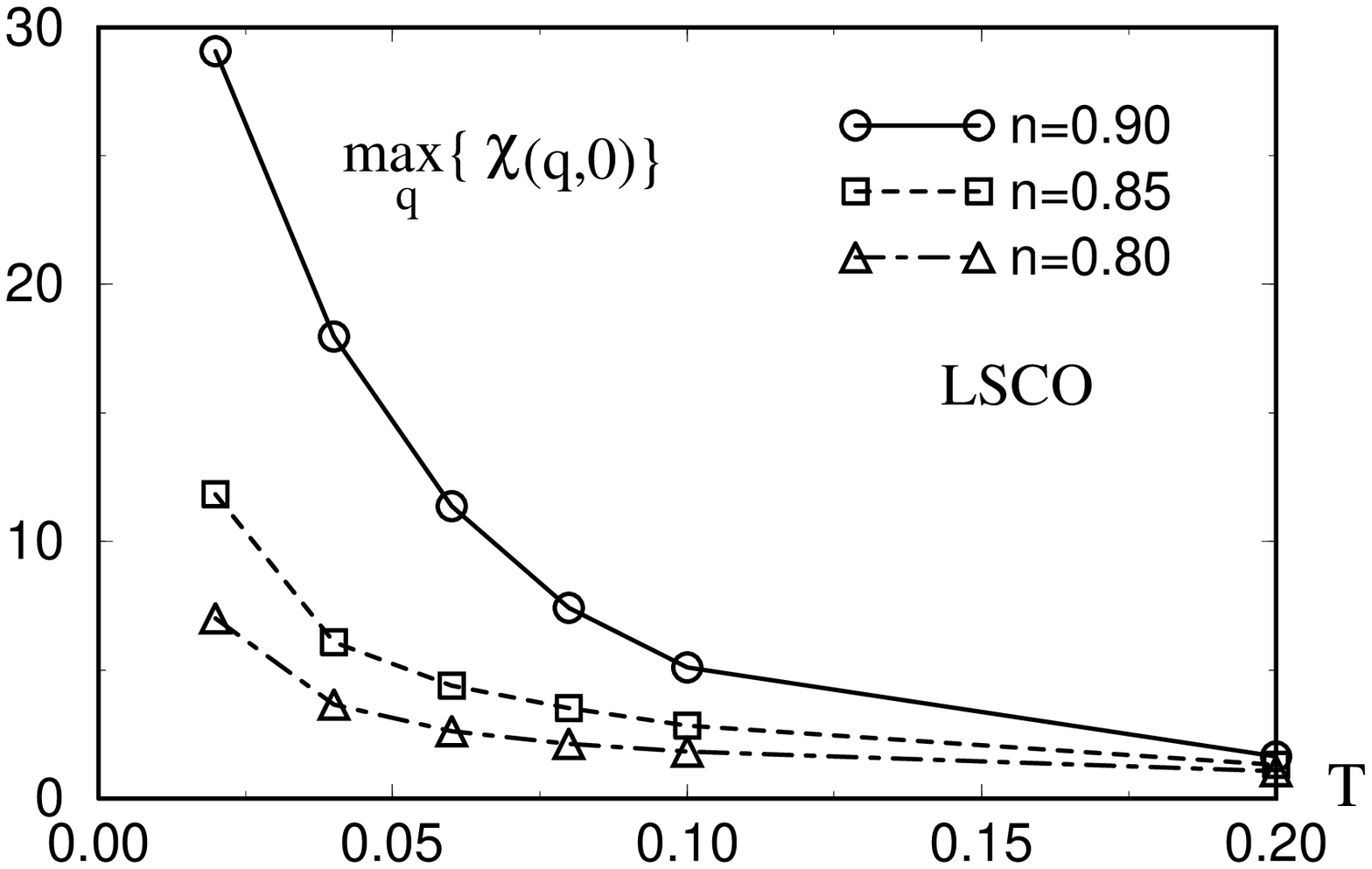}}
\caption{The temperature dependence of $\max_\q\{\chi(\q,0)\}$
 for various filling numbers.
 All of them follow the Curie-Weiss law, which is a universal
 feature of HTSC's.
 In YBCO and NCCO, $\chi_\q(0)$ takes the maximum value
 at $\q=\Q$. We note that $\chi_\Q(0)\propto\xi^2$.}
\label{fig:AF}
\end{figure}

Figure \ref{fig:AF} shows the temperature dependence of 
$\max_\q\{\chi_\q(0)\}$ of 
YBCO, NCCO and LSCO for different filling numbers.
These plots are nothing but the $T$-dependence of $\xi^2$.
Various experimental works on HTSC's by neutron diffraction
or by NMR confirm that 
$\xi^2$ follows the Curie-Weiss law qualitatively for $T>T^\ast$,
and its Curie constant increases as $n\rightarrow1$.
 \cite{Mag_scaling,pseudo_scaling}
As shown in Fig. \ref{fig:AF},
the FLEX approximation reproduce both the temperature and the doping 
dependence of $\xi^2$ in HTSC's for $T\simle0.1$.
However, the calculation of $\chi_\q(0)$ including the vertex corrections 
will be required for more detailed studies.

\begin{figure}
\epsfxsize=68mm
\centerline{\epsffile{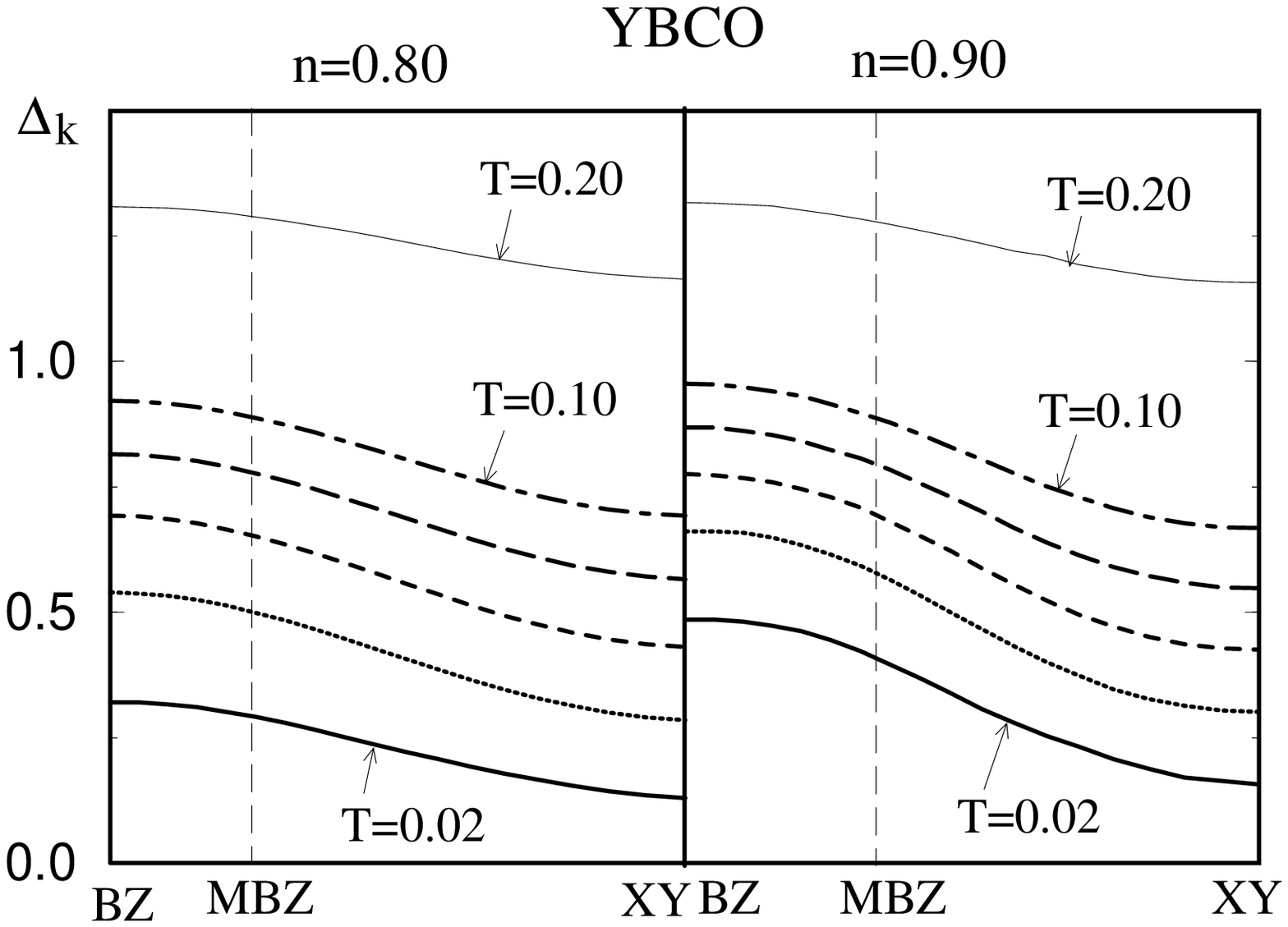}}
\epsfxsize=68mm
\centerline{\epsffile{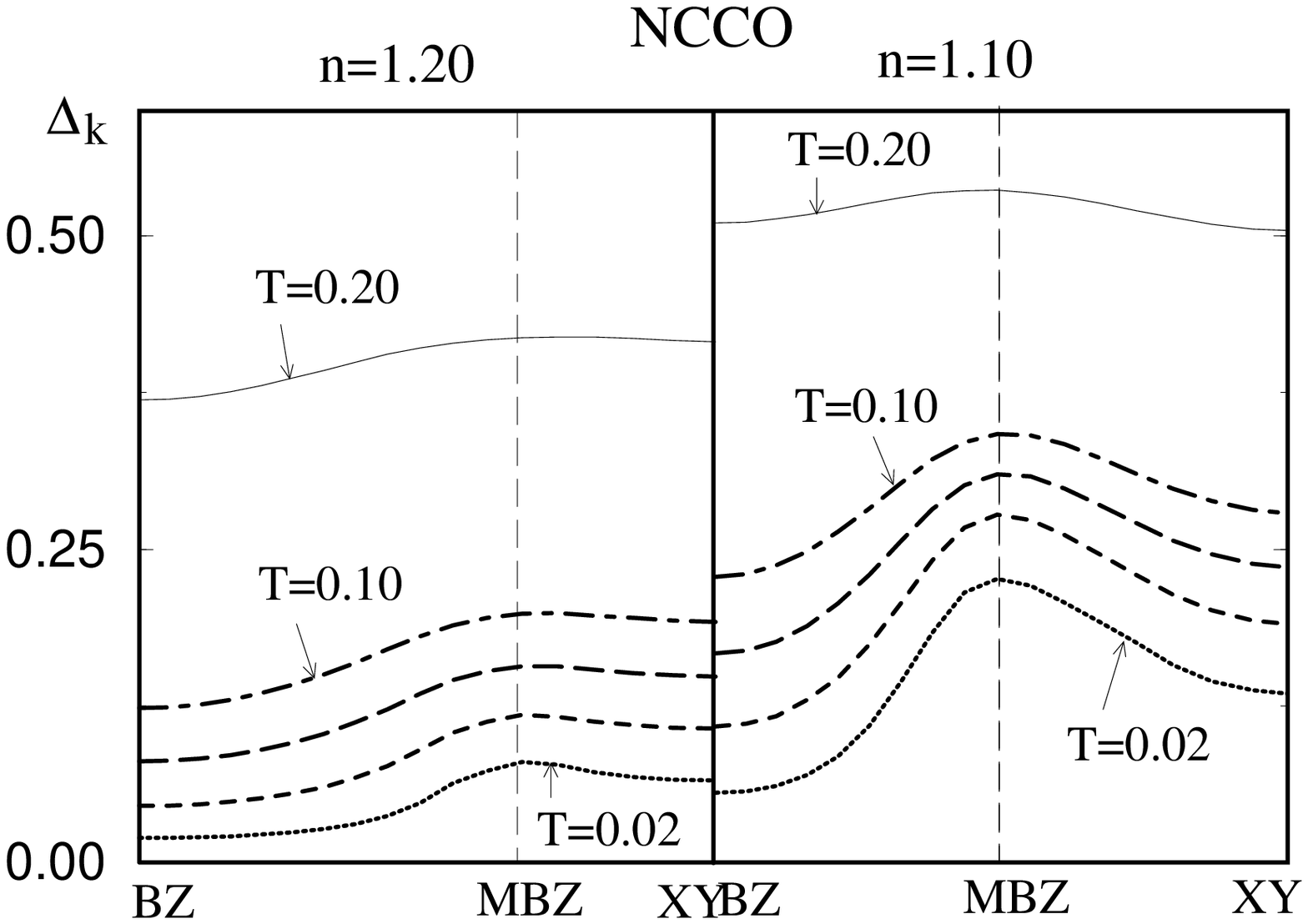}}
\epsfxsize=68mm
\centerline{\epsffile{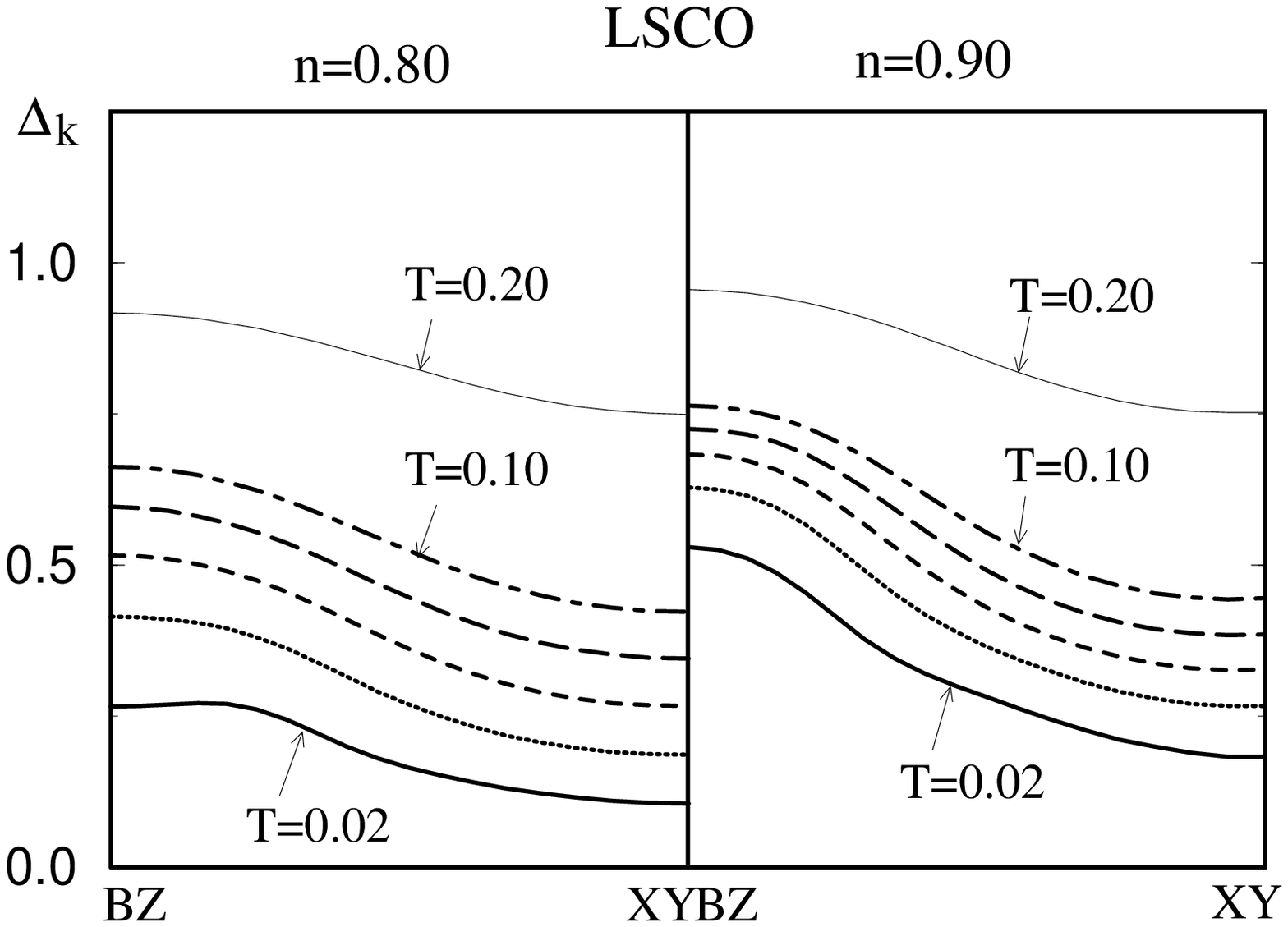}}
\caption{The $k_\parallel$-dependence of $\Delta(k_\parallel)$ 
 at various temperatures.
 The $T$-dependence of $\Delta(k_\parallel)$ at the cold spot
 and the hot spot are quite different.}
\label{fig:Del}
\end{figure}

\begin{figure}
\epsfxsize=50mm
\centerline{\epsffile{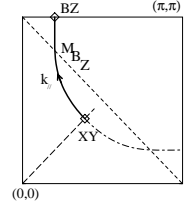}}
\caption{The path of $k_\parallel$ in the case of YBCO.}
\label{fig:path}
\end{figure}

Finally, we discuss the $\k$-dependence of
$\Delta_\k= {-\rm Im} \Sigma_\k(0+\i\delta)$ on the FS.
Here, we define
\begin{eqnarray}
\Delta(k_\parallel) \equiv \int dk_\perp \Delta_\k\cdot\rho_\k(0)
/ \int dk_\perp \rho_\k(0),
\end{eqnarray}
where $k_\parallel$ and $k_\perp$ are the momentum 
parallel and perpendicular to the FS, respectively.
$\Delta(k_\parallel)$ is an averaged value of $\Delta_\k$
over the $k_\perp$-direction on the FS, 
which 
has a finite width at finite temperatures.
Figure \ref{fig:Del} shows the $k_\parallel$-dependence of 
$\Delta(k_\parallel)$ over the 1/8-part of the FS, 
as shown in Fig. \ref{fig:path}. 
In each cases, the relation $z_\k\Delta_\k \sim T$ is realized
around the cold spots because $1/z_\k \simle 10$ is satisfied.


For YBCO and LSCO, $\Delta(k_\parallel)$ takes maximum value
not on the hot spot shown in Fig. \ref{fig:FS-hotcold},
but on the BZ-boundary,
because of the influence of the van-Hove singularity around $(\pi,0)$.
As a result, 
the spectral weight at the Fermi energy is reduced around $(\pi,0)$,
which is consistent with ARPES experiments.
And $\Delta(k_\parallel)$ takes minimum at the cold spot.
On the other hand, for NCCO,
the hot spot locates on the XY-axis,
and the cold spot locates on the BZ-boundary.
It is quite important that the position of the cold spots changes 
across $n\approx 1$ by using the FLEX approximation,
which causes the change of sign of $R_{\rm H}$ as shown in \S V.
To verify this result experimentally, 
the ARPES measurements for NCCO are desired.

As shown in Fig. \ref{fig:Del},
$\Delta(k_\parallel)$ at the cold spots decreases 
in proportion to $T$ qualitatively in all cases.
To understand its behavior, we calculate the following quantity:
\begin{eqnarray}
\langle\Delta_\k\rangle_{\rm FS} &\equiv& \sum_\k \Delta_\k \rho_\k(0)
 \nonumber \\
&=& \frac{3U^2}{4} \sum_\q 
 \frac{\chi_Q \w_{\rm sf} \cdot (\pi T)^2} {\w_\q(\w_\q + \pi T/2)} 
 \cdot \left.\frac{\d {\rm Im}\chi_\q^0(\w\!+\!\i\delta)}
  {\d\w}\right|_{\w=0}
  \nonumber \\
&\propto& \chi_Q T \xi^{-2} \cdot
 \left( 1-\left(1+ \pi T/2\w_{\rm sf} \right)^{-1/2} \right).
  \label{eqn:Delta_av}
\end{eqnarray}
We notice that the $\q$-dependence of
$\frac{\d}{\d\w}{\rm Im}\chi_\q^0(\w+\i\delta)|_{\w=0}$
in the FLEX approximation is moderate.
Here we assume $|\Q-(\k-\k')|\simle\xi^{-1}$ 
even if $\k$ and $\k'=\pm(k_y,k_x)$ are on the cold spots.
Then, $\langle\Delta_\k\rangle_{\rm FS}$
is expected to reflect $\Delta_{\rm cold}$
because $\rho_\k(0)$ takes smaller values around the hot spots.
As a result,
\begin{eqnarray}
\Delta_{\rm cold} \propto \left\{
\begin{array}{llll}
               \xi^2 T^2  & \propto T &
               \ \ \ \ {\rm for} \ \ \w_{\rm sf}\simge T  , \\
               \xi^0 T & \propto T &
               \ \ \ \ {\rm for} \ \ \w_{\rm sf}\simle T , 
\end{array}
\right.
   \label{eqn:Delta_av2}
\end{eqnarray}
which is independent of $\xi$ for $\w_{\rm sf}\simle T$ 
(under-doped region).
This means that 
$\rho \propto T$ is expected for wider range of filling $n$,
which is consistent with experiments.
In \S VI.B,
we calculate $\rho$ accurately according to the Kubo formula.

At the hot spots, however, $\Delta(k_\parallel)$ 
deviates from $T$-linear behavior.
In fact,
$\Delta(k_\parallel)$ at the hot spot is given by using 
(\ref{eqn:Delta}) as
\begin{eqnarray}
\Delta_{\rm hot}&=& \frac{3U^2}{4\pi} \int_{\rm FS}
 \frac{d q_{\parallel}}{|v_\q|}
 \chi_Q \w_{\rm sf} \frac{(\pi T)^2}{4\w_\q (\w_\q+\pi T/2)}
  \nonumber \\
&\approx& \frac{3\pi U^2}{2|v|} \chi_Q T \xi^{-1} \cdot
 \left(1-\left(1+\pi T/2\w_{\rm sf}\right)^{-1/2}\right) .
  \label{eqn:Delta_hot} 
\end{eqnarray}
As a result,
\begin{eqnarray}
\Delta_{\rm hot} \propto \left\{
\begin{array}{lll}
 T^2\xi^3 & \propto \sqrt{T} &
 \ \ \ \ {\rm for} \ \ \w_{\rm sf}\simge T , \\
 T\xi & \propto \sqrt{T}  &
 \ \ \ \ {\rm for} \ \ \w_{\rm sf}\simle T .
\end{array}
\right.
\end{eqnarray}
Thus, we find that $\Delta_{\rm hot} \propto \sqrt{T}$
for wider range of filling $n$, which seems to be realized
in Fig. \ref{fig:Del} in all the cases.
This relation does not contradict with the $T$-linear resistivity
because $\rho$ is determined mainly by the cold spot properties.

\begin{figure}
\epsfxsize=68mm
\centerline{\epsffile{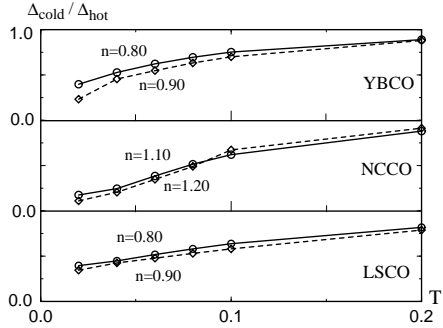}}
\caption{The temperature dependence of 
 $\Delta_{\rm cold}/\Delta_{\rm hot}$.
}
\label{fig:Ratio}
\end{figure}

Finally, we show the temperature dependence of the 
anisotropy of $\Delta(k_\parallel)$,
$r=\Delta_{\rm cold}/\Delta_{\rm hot}$
in Fig. \ref{fig:Ratio}.
In all cases, $r$ becomes smaller in the under-doped region,
which is consistent with recent ARPES experiments.
 \cite{ARPES-hot}
However, we see that $r$ depends on the shape of the FS's sensitively.
The relation $r \propto \sqrt{T}$, which is expected 
according to eqs. (\ref{eqn:Delta_av}) and (\ref{eqn:Delta_hot}),
is satisfied clearly only in YBCO.
In conclusion, the relation $r \propto \sqrt{T}$
is less universal than the Curie-Weiss behavior of
$\chi_Q$ in HTSC,
which is reproduced by the FLEX approximation for all compounds.
(see Fig. \ref{fig:AF}).

\subsection{Resistivity and Hall coefficient}
In \S V, we find that the vertex correction
from the MT term gives singular behaviors.
In this subsection, we obtain the self-consistent solution for
$\s_{xx}$ and $\s_{xy}/H$,
by using the self-consistent Green function obtained by 
the FLEX approximation.
We solve the BS equation for $J_{\k\mu}(\w)$ explicitly
by including all the MT terms.
Here we do not use eqs. (\ref{eqn:sigma_xx}), (\ref{eqn:def_Jx}) 
and (\ref{eqn:Kohno_Yamada2}) because 
the energy-integrations in deriving them have been done 
under the assumption that $z_k \Delta_\k \ll T$.
In reality, $z_k \Delta_\k \sim T$ is realized
in the present case.
Below, we perform the energy-integration seriously
by taking account of the energy dependence of 
$v_{\k\mu}(\w)=v_{\k\mu}^0+d{\rm Re}\Sigma_\k(\w)/dk_\mu$, 
$\Delta_\k(\w)={-\rm Im} \Sigma_\k(\w+\i\delta)$,
$V_k(\w)$, and $J_{\k\mu}(\w)$.

To obtain $\s_{xx}$ and $\s_{xy}/H$, 
we solve the following equations self-consistently:
\begin{eqnarray}
& &\sigma_{xx} = {e}^2 \sum_\k \int \frac{d\e}{\pi}
 \left(-\frac{\d f}{\d\e} \right) 
 \left( \ |G_\k(\e)|^2 \cdot v_{\k x}(\e) J_{\k x}(\e) \right.
  \nonumber \\
& & \ \ \ \ \ \ \ \ \ \ \ \ \ \ \ \ \ \ \ \ -  \left. 
 {\rm Re} \left\{G_\k^2(\e) \cdot v_{\k x}^2(\e) \right\} \ \right) ,
  \label{eqn:s_numerical} \\
& &\sigma_{xy}/H = -{e}^3 \sum_\k \int \frac{d\e}{2\pi}
 \left(-\frac{\d f}{\d\e} \right) |{\rm Im}G_\k(\e)| 
  \nonumber \\
& &\ \ \ \ \ \ \ \ \ \ \ \ \ \times
  |G_\k(\e)|^2 \cdot A_{s}(\k,\e), 
  \label{eqn:sH_numerical} \\
& &\ A_{s}(\k,\e) =
 v_{\k x}(\e) 
 \left[ J_{\k x}(\e) \frac{\d J_{\k y}(\e)}{\d k_y} 
       -J_{\k y}(\e) \frac{\d J_{\k x}(\e)}{\d k_y} \right] 
  \nonumber \\
& &\ \ \ \ \ \ \ \ \ \ \ \ \ + 
 \langle x \leftrightarrow y \rangle ,
  \label{eqn:A_numerical} \\
& &\ J_{\k \mu}(\w) = v_{\k \mu}(\w)+ \sum_{\q} \int \frac{d\e}{2\pi}
 \left[ {\rm cth}\frac{\e-\w}{2T} - {\rm th}\frac{\e}{2T} \right] 
 \nonumber \\
& &\ \ \ \ \ \ \ \ \ \ \  \ \ \ \times 
 {\rm Im}V_{\k-\q}(\e-\w+\i\delta) \cdot 
 |G_\q(\e)|^2 \cdot J_{\q \mu}(\e) ,
  \label{eqn:J_numerical} 
\end{eqnarray}
where $f(\e)=(\exp((\e-\mu)/T)+1)^{-1}$, and
$G(\w+\i\delta)$ and $\Sigma(\w+\i\delta)$ are derived 
from $G(\w_n)$ and $\Sigma(\w_n)$
through the numerical analytic continuation.
 \cite{Pade}
The $\e$-integration in the above equations are not difficult because
its leading contribution comes only from $|\e|\simle T$.
Note that $|G_\k(\w)|^2= \pi\rho_\k(\w)/\Delta_\k(\w)$
and $|G_\k(\w)|^2|{\rm Im}G_\k(\w)|
 \approx \pi\rho_\k(\w)/2\Delta_\k^2(\w)$.

As for the resistivity,
the existence of the second term of eq. (\ref{eqn:s_numerical}),
whose derivation will be publised elsewhere, 
has been overlooked in literatures so far.
It gives qualitatively an important contrivution
in the case of $z_\k \Delta_\k \sim T$, in where
the quasiparticles are not well-defined.

\begin{figure}
\epsfxsize=65mm
\centerline{\epsffile{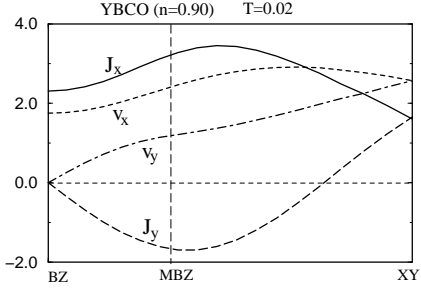}}
\caption{The obtained ${\vec J}(k_\parallel)$ 
together with ${\vec v}(k_\parallel)$ on the FS.}
\label{fig:DJ}
\end{figure}

Because $\Sigma_\k(\w)$ satisfies the self-consistency condition,
$v_{\k\mu}(\w)$ includes all the vertex corrections (a)-(c) 
in Fig. \ref{fig:vertex} automatically.
Although $J_\k(\w)$
contains only the (a) process of
${\cal T}_{\k\k'}(\e,\e')$ in the present calculation,
the others give only tiny corrections as shown in \S IV.
In this sense, our theory satisfies the condition of the 
conserving approximation well numerically.
We did not find any difficulty in solving
the BS equation (\ref{eqn:J_numerical})
for $J_{\k \mu}(\w)$ numerically,
since the self-consistency condition for $G_\k(\w)$
is satisfied in the FLEX approximation.
 \cite{Baym-Kadanoff,Baym}
Figure \ref{fig:DJ} shows the obtained ${\vec J}(k_\parallel)$
for YBCO on the FS along the path shown in Fig.\ref{fig:path}.
Its feature is close to the schematic one in Fig \ref{fig:J-schematic}.
We see that $J_y$ is negative around the hot spots in
this figure.
Apparently, such a region is enlarged in the case of NCCO.

\begin{figure}
\epsfxsize=70mm
\centerline{\epsffile{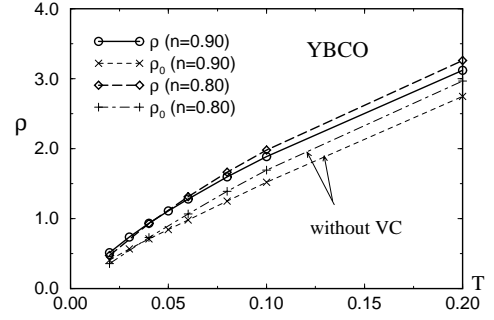}}
\epsfxsize=70mm
\centerline{\epsffile{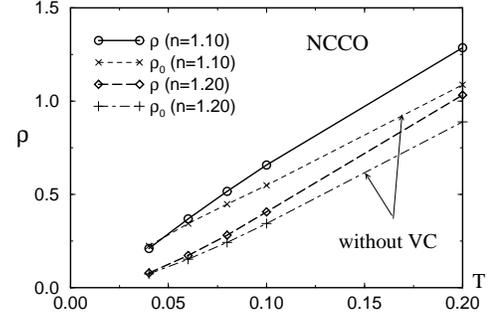}}
\epsfxsize=70mm
\centerline{\epsffile{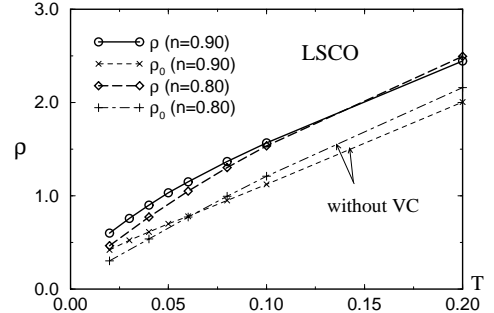}}
\caption{Temperature dependence of $\rho$ and $\rho^0$.
 We find $\rho>\rho^0$ in all the cases.
 Note that $\rho=1.0$ in this figure corresponds to 
 $\sim 4\times10^{-4}\Omega{\rm cm}$ in single layer compounds.
 'VC' means the vertex corrections for the current.
 We stress that $d\rho/dT$ increases below $T\approx0.08$ in
 YBCO and LSCO, while $d\rho^0/dT$ does not.
 This phenomena is caused by the VC, 
 not by the psuedo-gap formation in the DOS.
}
\label{fig:Rho}
\end{figure}

\begin{figure}
\epsfxsize=70mm
\centerline{\epsffile{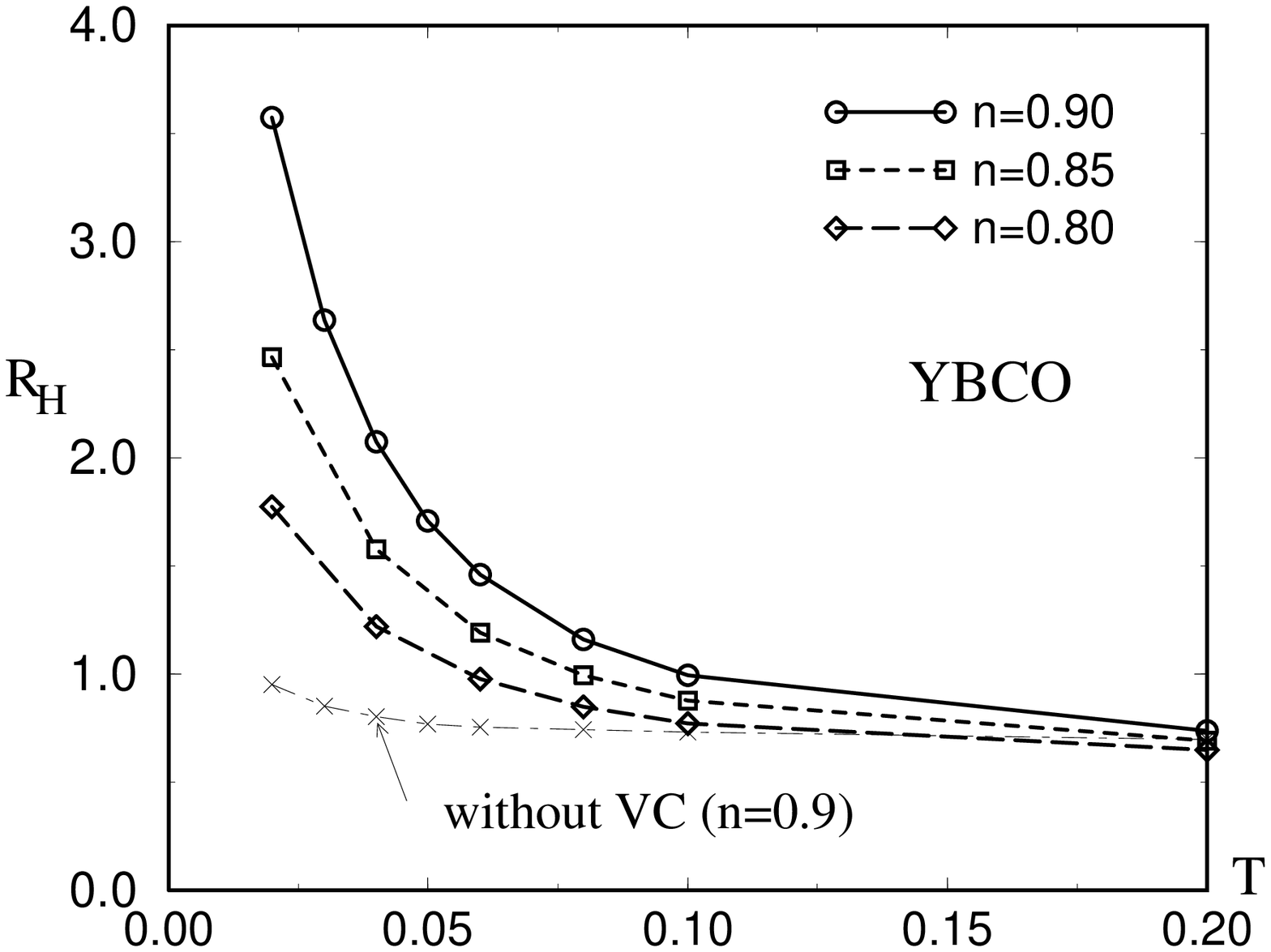}}
\epsfxsize=70mm
\centerline{\epsffile{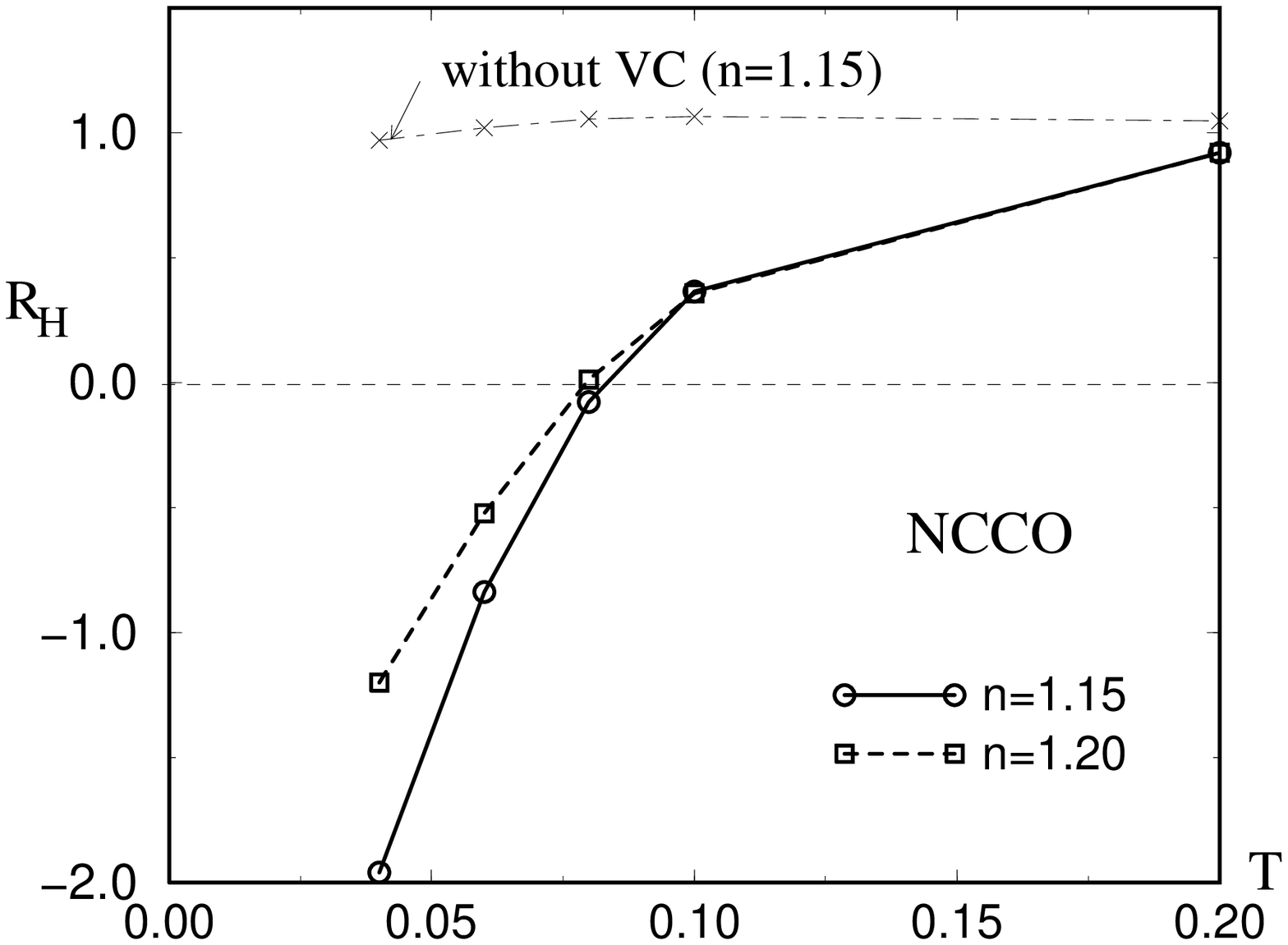}}
\epsfxsize=70mm
\centerline{\epsffile{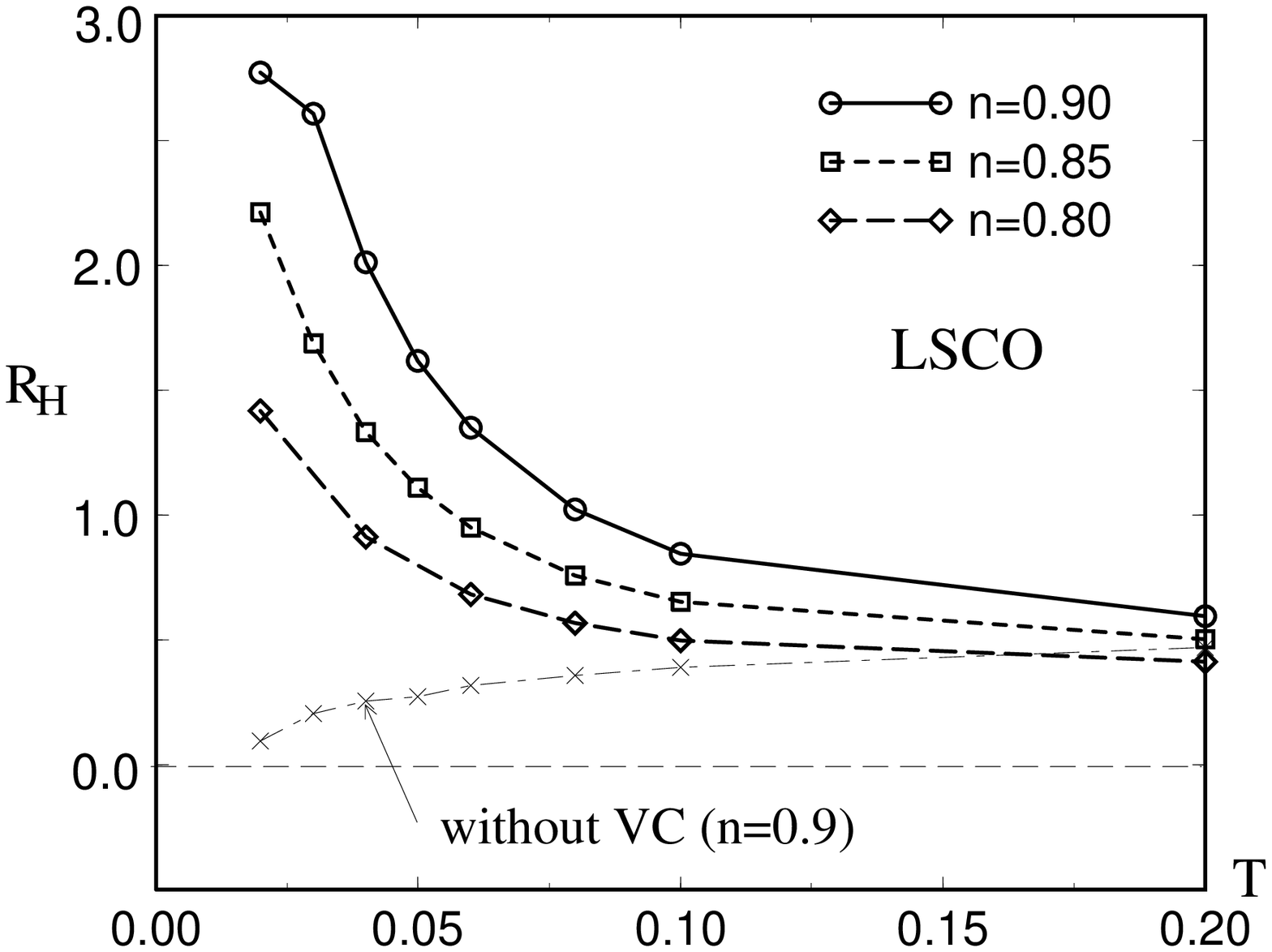}}
\caption{Temperature dependence of $R_{\rm H}$ and $R_{\rm H}^0$.
 We see that $R_{\rm H}$ (more precisely $R_{\rm H}-R_{\rm H}^0$)
 follows the Curie-Weiss type law in all the cases.
 This universal behavior is ascribed to the $T$-dependence of $\xi^2$.
 Here we put $e=1$.
 Note that $1/|ne|\sim 1.5\times10^{-3}{\rm cm}^3$/C in HTSC's.}
\label{fig:RH}
\end{figure}

Below, we examine the obtained numerical results for
YBCO, NCCO, and LSCO.
The calculated temperature dependence of 
$\rho=1/\s_{xx}$ and $R_{\rm H}=(\s_{xy}/H)\cdot\rho^2$ are shown
in Fig. \ref{fig:Rho} and Fig.\ref{fig:RH}, respectively.
In these figures, we also plot 
$\rho^0=1/\s_{xx}^0$ and $R_{\rm H}^0=(\s_{xy}^0/H)\cdot(\rho^0)^2$,
where $\s_{xx}^0$ and $\s_{xy}^0$ is given by 
replacing $J_{\k\mu}(\w)$ with $v_{\k\mu}(\w)$
in eqs. (\ref{eqn:s_numerical}) and (\ref{eqn:sH_numerical}).
Both $\s_{xy}^0$ and $\s_{xx}^0-\s_{\rm inc}$ are the result
of the conventional Boltzmann approximation,
where the conservation laws is violated.

{\underline{\it Resistivity :}} \\
At first, we discuss the $T$-dependences of the resistivity
shown in Fig. \ref{fig:Rho}, where we put $e^2/\hbar=1$.
$T=0.1$ corresponds to $\sim 500$K
if we assume $|t_0|\sim0.5$eV.
In every case, both $\rho^0$ and $\rho$ show an approximate 
$T$-linear behavior, 
reflecting the temperature dependence of $\Delta_\k$
at the cold spots as shown in Fig. \ref{fig:Del}.
 \cite{Hall-Pines,Yanase}
They are consistent with experiments.
In all the cases the relation $\rho>\rho^0$ is realized,
as is expected from the analysis in \S V.
In LSCO and YBCO, the extrapolated value of $\rho^0$ at $T=0$
from the higher temperature region is zero, 
while that of $\rho$ seems to take a finite value 
even in a pure system.
This behavior of $\rho$ can be explained 
by looking at eq. (\ref{eqn:vJ}) 
because $\a_\k$ decreases as $T$ increases,
reflecting the decrease of the 
backward scattering processes at higher temperatures.

The doping dependence of $\rho$ in YBCO and LSCO
is very small by using the present set of parameters.
According to Ref. \cite{Kimura}, however,
$(\rho(300K)-\rho(100K)) \times 10^{-4}$ $[\Omega {\rm cm}]$ in
La$_{2-\delta}$Sr$_\delta$CuO$_4$ is about 4.0, 2.5 and 1.8
for $\delta$= 0.11, 0.18 and 0.28, respectively.
This discrepancy with experiments is an important future problem, which
will be improved by choosing a more appropriate set of parameters.
Besides, we can not reproduce the doping dependence of the residual 
resistivity observed experimentally because we neglect 
the impurity effect. 
We note that $R_{\rm H}$ is more insensitive to parameters than $\rho$ is
because $R_{\rm H}$ is essentially independent of the life-time of the 
quasiparticles.

{\underline{\it Hall Coefficient :}} \\
Next, we discuss $R_{\rm H}$ shown in Fig. \ref{fig:RH}.
At higher temperatures ($T\sim 0.2$),
where $\xi < 1$ is satisfied, 
we see that $R_{\rm H}\approx R_{\rm H}^0$ for all compounds.
As the temperature decreases, $R_{\rm H}$ increases
for YBCO and LSCO, and decreases for NCCO,
and $|R_{\rm H}-R_{\rm H}^0|$ follows the Curie-Weiss like 
behavior in all cases.
Moreover, its coefficient increases rapidly as the filling 
approaches to $n=1$, which is consistent with the 
experimental relation $R_{\rm H}\propto |1-n|^{-1}$. 
These behaviors are consistent with the analysis in \S V.
The obtained filling dependence of $R_{\rm H}$ is much stronger than
that of $\rho$ (or $d\rho/dT$).

Moreover, the sign of $R_{\rm H}$ in NCCO changes to negative 
below $T\simle0.08\sim400$K,
which is consistent with experiments.
The Boltzmann approximation can not explain this behavior
because the shape of the FS is hole-like everywhere.
In the case of LSCO, 
the FS around $(\pm\pi,0)$ or $(0,\pm\pi)$ are convex
in the over-doped region.
In this case, the Hall coefficient can be negative
within the Boltzmann approximation.
Experimentally, $R_{\rm H}$ in La$_{1-x}$Sr$_x$CuO$_2$ becomes
negative and almost temperature independent for $x\simge0.32$, 
where no superconducting transition occurs and 
the AF fluctuations are much weaker.
In the present calculation for LSCO at $n=0.65$, 
we find that $R_{\rm H}$ is nearly zero for $T=0.2\sim0.02$, 
and $R_{\rm H}\approx R_{\rm H}^0$ is realized.
As a result, important features of $R_{\rm H}$ in LSCO are reproduced
in the present study for $|1-n|\ge 0.1$.

Here, we consider the $T$-dependence of the Hall coefficient
given by neglecting the vertex corrections, $R_{\rm H}^0$.
According to Fig. \ref{fig:RH},
$R_{\rm H}^0$ decreases moderately in LSCO and NCCO,
while $R_{\rm H}^0$ of YBCO slightly increases at $T<0.05$.
First, 
we discuss the effect of the $T$-dependence of 
$r=\Delta_{\rm cold}/\Delta_{\rm hot}$ on the Hall coefficient.
Figure \ref{fig:Ratio} shows that $r$ decreases as the temperature
decreases, and $r\propto\sqrt{T}$ is realized well for YBCO at $n=0.90$.
This effect makes $|R_{\rm H}^0|$ larger at low temperatures
because 
the effective number of carriers, which contributes to the transport
phenomena, decreases around the hot spots.
This mechanism has been pointed out by several authors 
to explain the enhancement of $R_{\rm H}^0$ in YBCO.
 \cite{Hall-Pines,Yanase}
Secondly,
we discuss the effect of the $T$-dependence of the FS.
As shown in Fig. \ref{fig:FS-T},
the curvature of the FS around the cold spots decreases 
as $T$ decreases, which should make $|R_{\rm H}^0|$ smaller.
In conclusion, 
through the cancellation of these two effects,
the Hall coefficient given by the Boltzmann transport approximation
is nearly $1/ne$ and is not enhanced significantly, if we take the 
temperature dependence of the FS into account correctly.

We stress that $R_{\rm H}$ in our calculation follows the 
Curie-Weiss law, even if the $T$-dependence of the FS is 
taken into account.
Undoubtedly, 
this behavior of $R_{\rm H}$ comes from the vertex corrections 
for the current.
In fact, in the present calculation,
the relation $R_{\rm H} \propto \chi_Q$
given by eq. (\ref{eqn:RH-est})
seems to be satisfied qualitatively 
for all the sets of parameters.
In summary, the vertex correction for the current is essential
for the Curie-Weiss behavior of $R_{\rm H}$ in high-T${\rm c}$ cuprates observed
experimentally.  This universal behavior of $R_{\rm H}$ is quite
robust in the present calculation.

\begin{figure}
\epsfxsize=75mm
\centerline{\epsffile{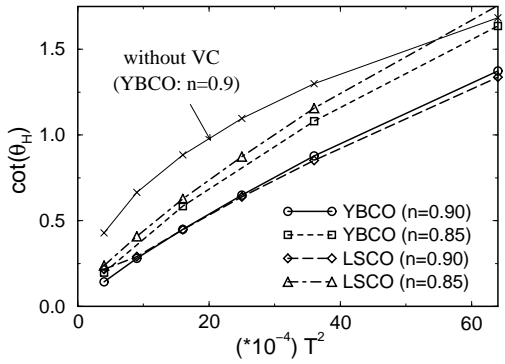}}
\caption{$T$-dependence of the Hall angle as 
 cot$(\Theta_{\rm H})$ vs $T^2$ in YBCO and LSCO,
 for $T=0.02\sim0.08$. (i.e., $T=120\sim480$K.)
 The thin line represents the cot$(\Theta_{\rm H}^0)$
 obtained within the Boltzmann approximation.}
\label{fig:HAng-T2}
\end{figure}

{\underline{\it Hall Angle :}} \\
We also discuss the temperature dependence of 
the Hall angle $\Theta_{\rm H}$, which is defined by
cot$(\Theta_{\rm H})= \s_{xx}/(\s_{xy}/H) = \rho/R_{\rm H}$.
Figure \ref{fig:HAng-T2} shows that 
cot$(\Theta_{\rm H})$ is approximately proportional to $T^2$
for $0.02\le T\le0.08$.
This relation has been observed experimentally
in various kinds of HTSC's for $T=100\sim300$K.
 \cite{Takagi,Peng,Chien}
Our theory can explain this relation 
without assuming the non-Fermi liquid ground state
which possesses two kinds of relaxation rates.
 \cite{Anderson}
Figure \ref{fig:HAng-T2} means that $R_{\rm H}$ follows the 
Curie-Weiss behavior, because $\rho$ is proportional to $T$.
We stress that the relation cot$(\Theta_{\rm H})\propto T^2$
is also observed experimentally in $\kappa$-BEDT-TTF compounds
 \cite{Sushko}
or in V$_2$O$_3$,
 \cite{V2O3-2}
both of them are also nearly AF-Fermi liquids.

Here, we discuss the momentum dependence of the following functions:
\begin{eqnarray}
& &{S}_{xy}(k_\parallel) \equiv 
 -\int dk_{\perp} \rho_\k(0) A_{s}(\k,0) \frac{1}{\{\Delta_\k(0)\}^2}
 \nonumber \\
& &\ \ \ \
 = -|{\vec J}(k_{\parallel})|^2
 \left(\frac{d\theta_J(k_{\parallel})}{dk_{\parallel}} \right)
 \frac{1}{\{\Delta(k_\parallel)\}^2}
\end{eqnarray}
where $k_\parallel$ is the momentum along the FS.
$A_{s}(\k,\e)$ is given by eq. (\ref{eqn:A_numerical}),
and $k_\perp$ is the momentum perpendicular to the FS.
It is clear that
$\s_{xy}/H \propto \int_{\rm FS}dk_\parallel S_{xy}(k_\parallel)$.
We also define ${S}_{xy}^0(k_\parallel)$ by replacing $J_{\k\mu}$
with $v_{\k\mu}$ in eq. (\ref{eqn:A_numerical}),
which means that 
$\s_{xy}^0/H \propto \int_{\rm FS}dk_\parallel S_{xy}^0(k_\parallel)$.

\begin{figure}
\epsfxsize=65mm
\centerline{\epsffile{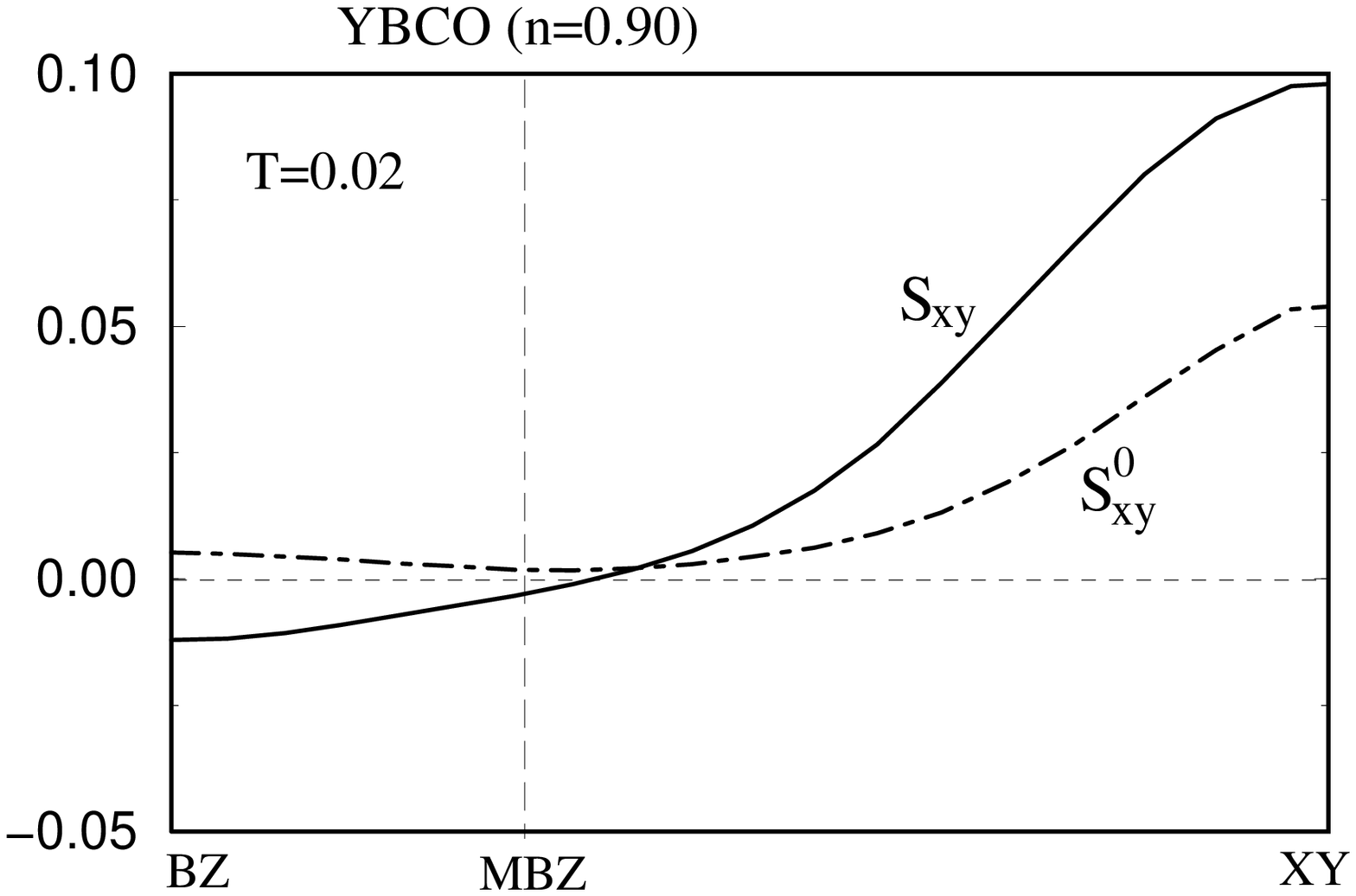}}
\epsfxsize=65mm
\centerline{\epsffile{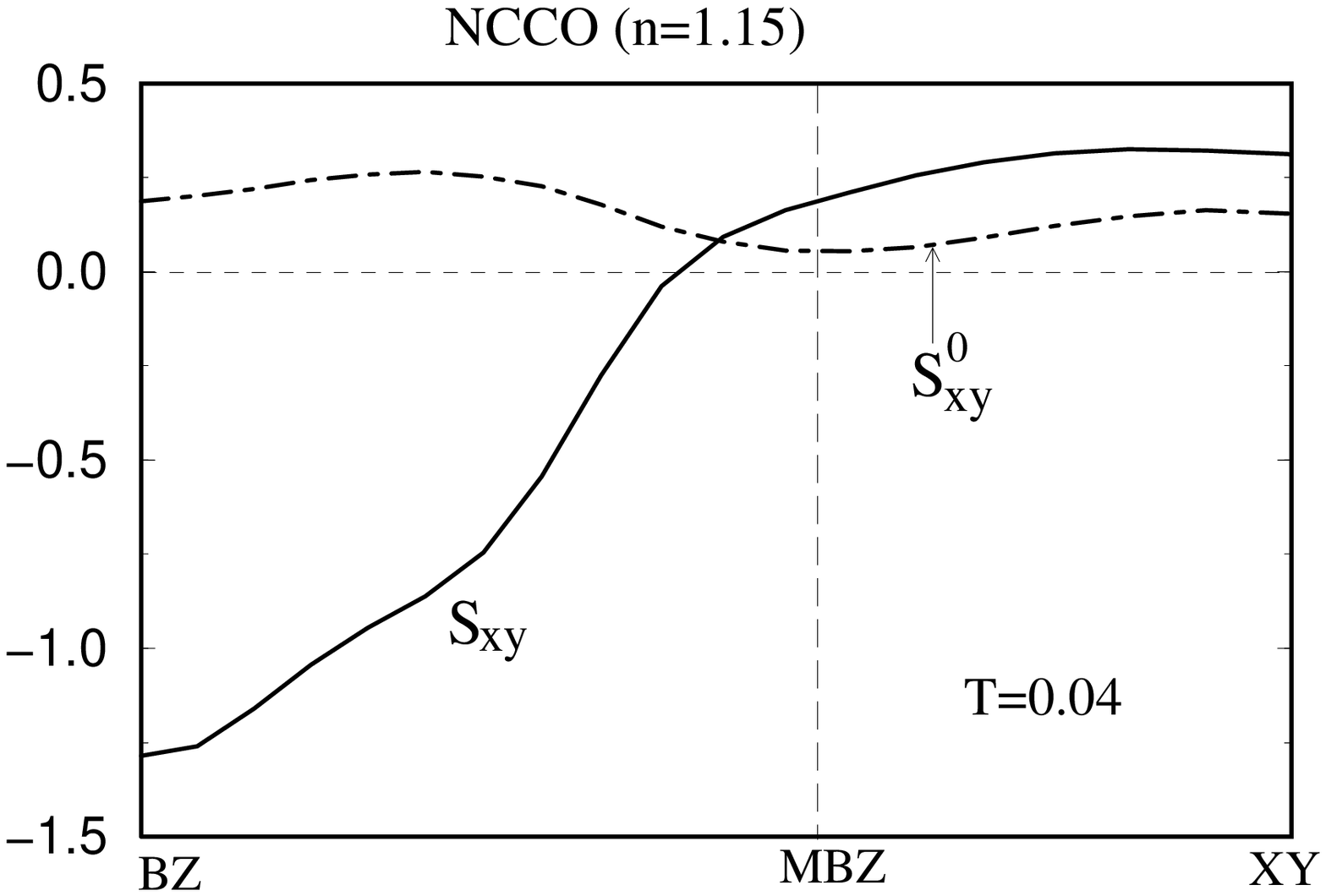}}
\caption{The momentum dependence of $S_{xy}(\k)$ 
 and $S_{xy}^0(\k)$ on the FS.
 } (see Fig. \ref{fig:path}.)
\label{fig:DHall}
\end{figure}

Figure \ref{fig:DHall} shows the momentum dependence
of $S_{xy}(k_\parallel)$ and $S_{xy}^0(k_\parallel)$
along the FS shown in Fig. \ref{fig:path}.
In both cases of YBCO and NCCO, $S_{xy}^0(k_\parallel)$
is positive everywhere. 
Whereas, $S_{xy}(k_\parallel)$ is positive inside the MBZ and
negative outside of it, which is consistent with the
analysis in \S V.

In the case of YBCO,
$S_{xy}(k_\parallel)$ takes a maximum value on the XY-axis
i.e., on the cold spot.
It takes an enhanced value because the relation
$S_{xy}(k_{\rm cold})/S_{xy}^0(k_{\rm cold})
 =1/(1-\a^2(k_{\rm cold})) \propto \xi^2$
is expected according to eq. (\ref{eqn:vJJ}).
As a result, $R_{\rm H}\propto \xi^2$ is realized.
We have also calculated $S_{xy}(k_\parallel)$ for LSCO,
and found that its behavior is similar to that for YBCO
in spite of the incommensurability of $\chi(q,0)$.
On the other hand, in the case of NCCO,
$S_{xy}(k_\parallel)$ takes a maximum value on the BZ-boundary
which is a cold spot of NCCO.
It is also enhanced because
$S_{xy}(k_{\rm cold})/S_{xy}^0(k_{\rm cold})
 \propto \beta_\parallel(k_{\rm cold}) \propto \xi^2$ 
is expected by eq.(\ref{eqn:vJJ}).
As a result, $R_{\rm H}\propto -\xi^2$ is realized and
$R_{\rm H}$ becomes negative at low temperatures.

\section{Summary and Discussions}

First, we outline the main results of this paper:
We have calculated the conductivity $\s_{xx}$ and the Hall conductivity 
$\s_{xy}/H$ in the single-band Hubbard model
based on the Fermi liquid theory.
We have calculated the total current ${\vec J}_{\k}$ including 
the vertex corrections in the conservation approximation.
In nearly AF Fermi liquids, 
the Bethe-Salpeter equation (\ref{eqn:def_Jx}) for ${\vec J}_{\k}$
can be simplified to eq. (\ref{eqn:BS_simple}) approximately.
The obtained ${\vec J}_{\k}$ 
shows non-trivial critical behaviors as seen in Fig. \ref{fig:J-schematic},
which is the natural consequence of the strong backward scatterings.
In conclusion, $R_{\rm H}\propto \xi^2$ is realized in HTSC's
through the anomaly of ${\vec J}_{\k}$.
This mechanism 
has not been pointed out previously.

We also have done the numerical calculations by using the FLEX approximation.
We can reproduce characteristic features of the spin
fluctuations for YBCO, NCCO and LSCO,
by using the appropriate set of parameters.
In each cases, $d_{x^2-y^2}$ superconductivity is realized at 
$T_{\rm c}=50\sim100$K.
Next, we have determined ${\vec J}_{\k}$ by solving eq.(\ref{eqn:def_Jx}) 
numerically, and calculated both $\rho$ and $R_{\rm H}$ for various 
filling numbers.
As shown in Fig. \ref{fig:RH}, the overall features of $R_{\rm H}$
in each compounds are reproduced quite well.
Especially, both the relations $R_{\rm H}\propto 1/T$ and $\rho\propto T$
are obtained at the same time.
We have found that $R_{\rm H}<0$ is realized in NCCO
because the cold spots in NCCO locate around the BZ-boundaries,
which may be verified by ARPES experiments.

The vertex corrections mentioned above 
are not included in the Boltzmann approximation.
We have confirmed that the Hall coefficient given by
the Boltzmann approximation $R_{\rm H}^0$ remains of order $O(1/ne)$
if we take the $T$-dependence of the FS into account correctly.
(see Fig.\ref{fig:RH}).
Moreover, $R_{\rm H}^0$ remains positive because the FS is
hole-like everywhere.
In conclusion, the anomalous behaviors of $R_{\rm H}$ in HTSC
is reproducible only if the vertex corrections for the current
are taken into account.

Here, we discuss the validity of the relation $R_{\rm H} \propto \xi^2$.
In a conserving approximation (including the FLEX approximation)
the interaction $V_\k(\w)$ which gives Im$\Sigma_\k(\w)$
also determine the MT-type vertex corrections for ${\vec J}_\k$,
as shown by eqs. (\ref{eqn:Delta}) and (\ref{eqn:JJJ}).
This condition leads to $\a_\k \sim 1$ in eq. (\ref{eqn:BS_simple}),
which strongly suggests that the relation $R_{\rm H} \propto \xi^2$ will be 
valid beyond the FLEX approximation.
Now we assume that it is valid 
near the half-filling case ($n\approx 1$).
Then, the experimental relation
$\max\{|R_{\rm H}|\} \propto 1/|1-n|$ can be understood
because $\max\{\xi^2\} \propto 1/|1-n|$ is observed experimentally
near the half-filling.
Next, we consider the Hall coefficient below 
the pseudo spin-gap temperature, $T^\ast$,
where $R_{\rm H}$ in YBCO decreases as $T$ decreases experimentally.
It is also understandable because 
$\xi$ slightly decreases below $T^\ast$ experimentally.
 \cite{Hg}

We also find that the resistivity is slightly enhanced 
by the vertex corrections for the current.
Moreover, we obtain the expression of the incoherent conductivity
$\s_{\rm inc}$, which is given by the second term of
eq.(\ref{eqn:s_numerical}).
In the case of $z_\k\Delta_\k \sim T$, which is satisfied in HTSC's, 
$\s_{\rm inc}$ $(>0)$ can be the same order of 
the first term of eq.(\ref{eqn:s_numerical})
at higher temperatures.

Unfortunately, the FLEX approximation 
becomes insufficient near the Mott-insulating state.
By this reason, we did not apply the present method
for $0.9\le n \le 1.1$.
Experimentally, both $|R_{\rm H}|$ and $d\rho/dT$ for $0.9\le n \le 1.1$
increases rapidly as $n\rightarrow 1$.
The FLEX approximation is also inappropriate for the study of 
electronic states below $T^\ast$, 
which is one of the important future problems on HTSC.
Recently, Ref.\cite{Pines-pseudogap} calculated some
vertex parts of the self-energy.
As an alternative possibility, the preformed pairs
may be formed for $T<T^\ast$.
This scenario has been intensively studied recently.
 \cite{Koikegami2,Dahm2}

In Ref. \cite{Kanki}, it is shown that 
a similar numerical study based on the AF spin-fluctuation model
also leads to the enhancement of $R_{\rm H}$,
by using the set of spin fluctuation parameters 
consistent with experiments.
Although the conserving laws are not satisfied exactly in that study,
it confirms the importance of the vertex corrections 
for the current.
It also indicates that the numerical results presented in this paper 
should not be taken as an artifact specific to the FLEX approximation.

Now we would like to discuss $R_{\rm H}$ 
in heavy Fermion (HF) compounds.
In various paramagnetic compounds, $R_{\rm H}$ shows a drastic
temperature dependence and takes an enhanced value.
 \cite{Onuki,Onuki2,Onuki3}
At low temperatures, the relation $R_{\rm H}=c\cdot\rho^2$ is
observed in many compounds, and $c$ is always positive.
It is explained in terms of the anomalous Hall effect (AHE), 
which originates from the localized $f$-orbital angular momentum,
and its enhancement factor is given by $\chi_0 \equiv \chi_{\q=0}(0)$.
 \cite{AHE-theory,AHE-theory2}
On the other hand, in some HF compounds with AF ground state,
the relation $R_{\rm H}=c\cdot\rho^2$ $(c>0)$
is not satisfied and the sign of $dR_{\rm H}/dT$ at $T>T_{\rm N}$
changes depending on compounds.
 \cite{Onuki,Onuki2,Onuki3}
Similar behavior is also observed in non-Fermi liquid HF compounds,
which is  near the AF quantum phase boundary,
e.g., Ce(Ni$_{1-x}$Pd$_x$)$_2$Ge$_2$.
 \cite{CeNiGe}
We stress that the normal Hall effect can exceed the AHE
and $R_{\rm H}\propto \chi_Q$ will be realized
in these nearly antiferromagnetic HF compounds,
where $\chi_Q \gg \chi_0$ is realized.
We note that $\chi_Q\propto(T-T_{\rm N})^{-3/2}$
in three dimension is obtained by the SCR theory.

We also comment on the $\kappa$-BEDT-TTF organic superconductors.
The recent studies by the FLEX approximation reveal
that the large AF fluctuations are the origin of the 
$d$-wave superconductivity.
 \cite{Kino,Kondo}
Recent experiments show that 
the $R_{\rm H}$ of this system increases as $T$ decreases
 \cite{Murata,Sushko},
which can be reproduced well by using the analysis of this paper.
 \cite{Kontani-BEDT}
Also, the relation $R_{\rm H}\propto 1/T$ is
observed in the superconducting ladder compound
Sr$_{14-x}$Ca$_x$Cu$_{24}$O$_{41}$.
 \cite{SCCO-RH},
whose characteristic electronic properties 
are well explained by the FLEX approximation.
 \cite{Trellis}
Moreover, $R_{\rm H}$ in V$_2$O$_3$ shows the 
singular $T$-dependence near the AF phase boundary.
 \cite{V2O3,V2O3-2}

\section*{Acknowledgments}
We are grateful to Kosaku Yamada for stimulating discussions.
We also thank T. Moriya, H. Fukuyama, M. Satoh, H. Kohno and Y. Yanase 
for valuable comments.
This work is financially supported by a Grant-in-Aid for Scientific
Research on Priority Areas from the Ministry of Education, 
Science, Sports and Culture.

\appendix
\section{Analytic Continuation for $\Gamma(\e_n,\e_{n'})$}

In this appendix, 
we derive the irreducible vertex ${\cal T}_{\k\k'}(\e,\e')$
which are the kernel of the BS equation, eq (\ref{eqn:def_Jx}).
For this purpose, we perform the analytic continuation
for $\Gamma(\e_n,\e_{n'};\w_l)$, where
$\w_l = 2\pi Tl$ ($l>0$) is the external frequency.
The irreducible vertices
consistent with the FLEX approximation 
are given by (\ref{eqn:A1})-(\ref{eqn:A3}).

\begin{figure}
\epsfxsize=40mm
\centerline{\epsffile{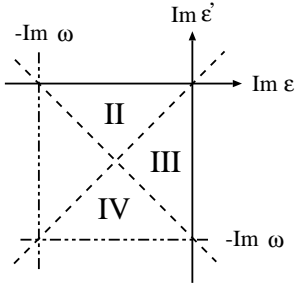}}
\caption{$\Gamma(\e,\e';\w)$ $(\w>0)$ is an analytic function 
inside of each (II,III,IV)-region, and has the cuts on each lines.}
\label{fig:region22}
\end{figure}

According to eq. (12) in ref. 
 \cite{Eliashberg},
\begin{eqnarray}
{\cal T}_{\k\k'}(\e,\e') &=& 
   {\rm cth}\frac{\e'-\e}{2T} 
  (\Gamma_{\k\k'}^{\II}(\e,\e')-\Gamma_{\k\k'}^{\III}(\e,\e'))
    \nonumber \\
& & +{\rm cth}\frac{\e'+\e}{2T} 
  (\Gamma_{\k\k'}^{\III}(\e,\e')-\Gamma_{\k\k'}^{\IV}(\e,\e'))
  \nonumber \\
& &  -{\rm th}\frac{\e'}{2T} 
  (\Gamma_{\k\k'}^{\II}(\e,\e')-\Gamma_{\k\k'}^{\IV}(\e,\e')).
 \label{eqn:T-def}
\end{eqnarray}
Here, 
$\Gamma^{\II}(\e,\e')$, $\Gamma^{\III}(\e,\e')$ and
$\Gamma^{\IV}(\e,\e')$  are given by the analytic continuations of
$\Gamma(\e_n,\e_{n'};\w_l)$
for regions II, III and IV in the complex $(\e,\e')$-plane
shown in Fig. \ref{fig:region22}, respectively.

Next, we take the limit $\i\w_l \rightarrow +0$.
For eq. (\ref{eqn:A1}), we get
\begin{eqnarray}
\Gamma_{\k\k'}^{(a)\II}(\e,\e')&=& V^R(\k-\k',\e'-\e), \\
\Gamma_{\k\k'}^{(a)\III}(\e,\e')&=&\Gamma_{\k\k'}^{(a)\IV}(\e,\e')
 \nonumber \\
 &=& V^A(\k-\k',\e'-\e),
\end{eqnarray}
where $A(R)$ represents the advanced (retarded) function.
Taking account of the relation Im$\{\e'-\e\}>0$ in the II-region
and Im$\{\e'-\e\}<0$ in the III,IV-region,
we get for eq. (\ref{eqn:A2}) as 
\begin{eqnarray}
& &\Gamma_{\k\k'}^{(b)\II}(\e,\e')
 = \sum_\q \int\frac{d\w}{2\pi} W_\q(\w) 
  \nonumber \\
& &\ \ \ 
 \times \left[ {\rm th}\frac{\w+\e}{2T} 
 {\rm Im}G_{\k+\q}^R(\w+\e) G_{\k'+\q}^R(\w+\e') \right.
 \nonumber \\
& &\ \ \ 
 + \left. {\rm th}\frac{\w+\e'}{2T} G_{\k+\q}^A(\w+\e) 
  {\rm Im}G_{\k'+\q}^R(\w+\e') \right]
 \nonumber \\
& &\ \ \ + C , \\
& &\Gamma_{\k\k'}^{(b)\III}(\e,\e')=\Gamma_{\k\k'}^{(b)\IV}(\e,\e')
 = \sum_\q \int\frac{d\w}{2\pi} W_\q(\w) 
  \nonumber \\
& &\ \ \ 
 \times \left[{\rm th}\frac{\w+\e}{2T} 
 {\rm Im}G_{\k+\q}^R(\w+\e) G_{\k'+\q}^A(\w+\e') \right.
 \nonumber \\
& &\ \ \
 + \left. {\rm th}\frac{\w+\e'}{2T} G_{\k+\q}^R(\w+\e) 
  {\rm Im}G_{\k'+\q}^R(\w+\e') \right]
 \nonumber \\
& &\ \ \ + C ,
\end{eqnarray}
where $C$ is a real function, and $W_\q(\w)$ is given by
\begin{eqnarray}
W_\q(\w)
   &=& \frac32 U^2 \left| U\chi_\q^{s}(\w)+1 \right|^2 
 \nonumber \\
   &+& \frac12 U^2 \left| U\chi_\q^{c}(\w)-1 \right|^2
   - U^2.
 \label{eqn:W-def2}
\end{eqnarray}
In the similar way, 
we can obtain the expressions of 
$\Gamma_{\k\k'}^{(c)\II}(\e,\e')=\Gamma_{\k\k'}^{(c)\III}(\e,\e')$ 
and $\Gamma_{\k\k'}^{(c)\IV}(\e,\e')$
through the analytic continuation of eq. (\ref{eqn:A3}), 
by taking account of 
the relations Im$\{\e'+\e+\w\}>0$ in the II,III-region and 
Im$\{\e'+\e+\w\}<0$ in the IV-region.

By inserting the above equations into eq. (\ref{eqn:T-def}),
${\cal T}_{\k,\k'}(\e_n,\e_{n'};\w_l)$ 
is given by ${\cal T}^{(a)}+{\cal T}^{(b)}+{\cal T}^{(c)}$.
They are derived as
\begin{eqnarray}
{\cal T}_{\k,\k'}^{(a)}(\e,\e') &=& 
 \left( {\rm cth}\frac{\e'-\e}{2T}-{\rm th}\frac{\e'}{2T} \right)
 \nonumber \\
& &\times 2\i \ {\rm Im}V_{\k'-\k}(\e'-\e+\i\delta), 
  \label{eqn:Ta} \\
{\cal T}_{\k,\k'}^{(b)}(\e,\e') &=&  
 \left( {\rm cth}\frac{\e'-\e}{2T}-{\rm th}\frac{\e'}{2T} \right)
 \sum_\q \int d\w W_\q(\w)
  \nonumber \\
& &\times (-\i\pi) 
 \left( {\rm th}\frac{\w+\e}{2T}-{\rm th}\frac{\w+\e'}{2T} \right)
 \nonumber \\
& &\ \times \rho_{\k+\q}(\w+\e) \rho_{\k'+\q}(\w+\e') ,
  \label{eqn:Tb} \\
{\cal T}_{\k,\k'}^{(c)}(\e,\e') &=& 
 \left( {\rm cth}\frac{\e'+\e}{2T}-{\rm th}\frac{\e'}{2T} \right)
 \sum_\q \int d\w  W_\q(\w)
  \nonumber \\
& &\times (-\i\pi) 
 \left( {\rm th}\frac{\w+\e}{2T}+{\rm th}\frac{\w-\e'}{2T} \right)
 \nonumber \\
& &\ \times \rho_{\k+\q}(\w+\e) \rho_{\k'-\q}(-\w+\e') ,
  \label{eqn:Tc}
\end{eqnarray}
Note that ${\cal T}_{\k\k'}^{(a{\mbox{-}}c)}(\e,\e')$
is pure imaginary.

\section{Corrections from the AL-terms for $J_{\k\mu}$}
In this appendix, we study the contributions from 
the AL-terms to the total current ${\vec J}_{\k}$.
For this purpose,
we solve the BS equation for ${\vec J}_{\k}(\w)$ 
including both the MT process and AL precesses,
and compare the obtained results with those in \S VI B.
The exact Bethe-Salpeter equation is,
\begin{eqnarray}
& &\ J_{\k \mu}(\w) = v_{\k \mu}(\w)+ \sum_{r}^{a,b,c}
 {\mit\Delta}J_{\k\mu}^{r}(\w) , 
 \label{eqn:J_numerical_all} \\
& &\ {\mit\Delta}J_{\k\mu}^{r}(\w) = \sum_{\q} \int_{-\infty}^{\infty}
 \frac{d\e}{4\pi\i} {\cal T}_{\k,\q}^{(r)}(\w,\e) \cdot
 |G_{\q}(\e)|^2 \cdot J_{\q \mu}(\e) ,
\end{eqnarray}
where 
$r=a,b,c$ and ${\cal T}_{\k,\q}^{(r)}(\w,\e)$
are given by eqs. (\ref{eqn:Ta})-(\ref{eqn:Tc}).
Note that ${\cal T}_{\k,\q}^{(r)}(\w,\e)$ are pure imaginary.

For simplicity of the numerical calculation,
we put all the energy variables in $\rho_{\k'}(\w)$,
$\Delta_{\k}(\w)$ and $J_{\k\mu}(\w)$ as zero
for ${\mit\Delta}J_{\k\mu}^{b}(\w)$ and ${\mit\Delta}J_{\k\mu}^{c}(\w)$.
Strict justification of this simplification is difficult,
although it may be sufficient for a rough estimation of the
magnitude of the AL terms.
We represent the solution of (\ref{eqn:J_numerical_all}) as
$J_{\k \mu}^{\rm MT+AL}(\w)$.

\begin{figure}
\epsfxsize=60mm
\centerline{\epsffile{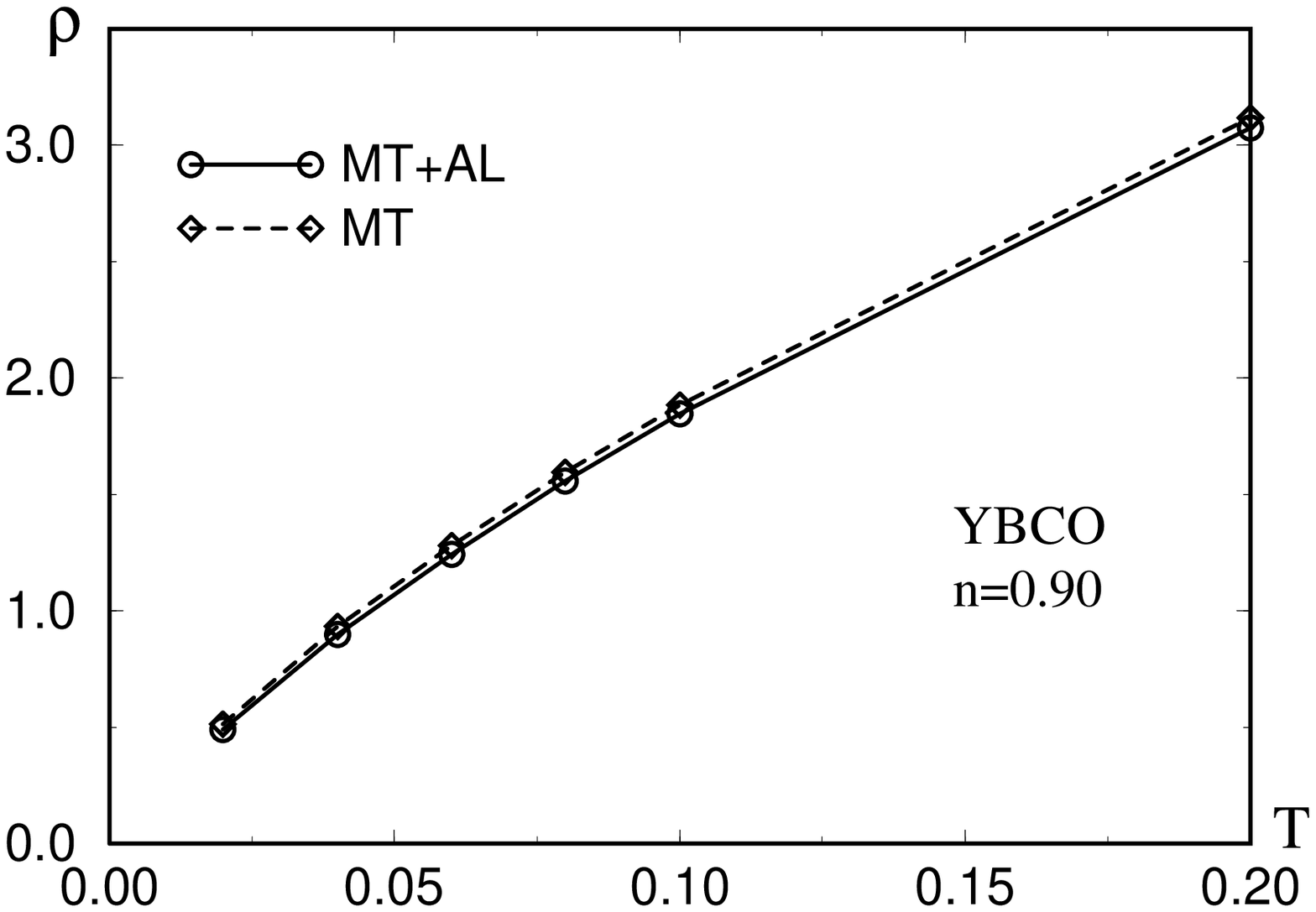}}
\epsfxsize=60mm
\centerline{\epsffile{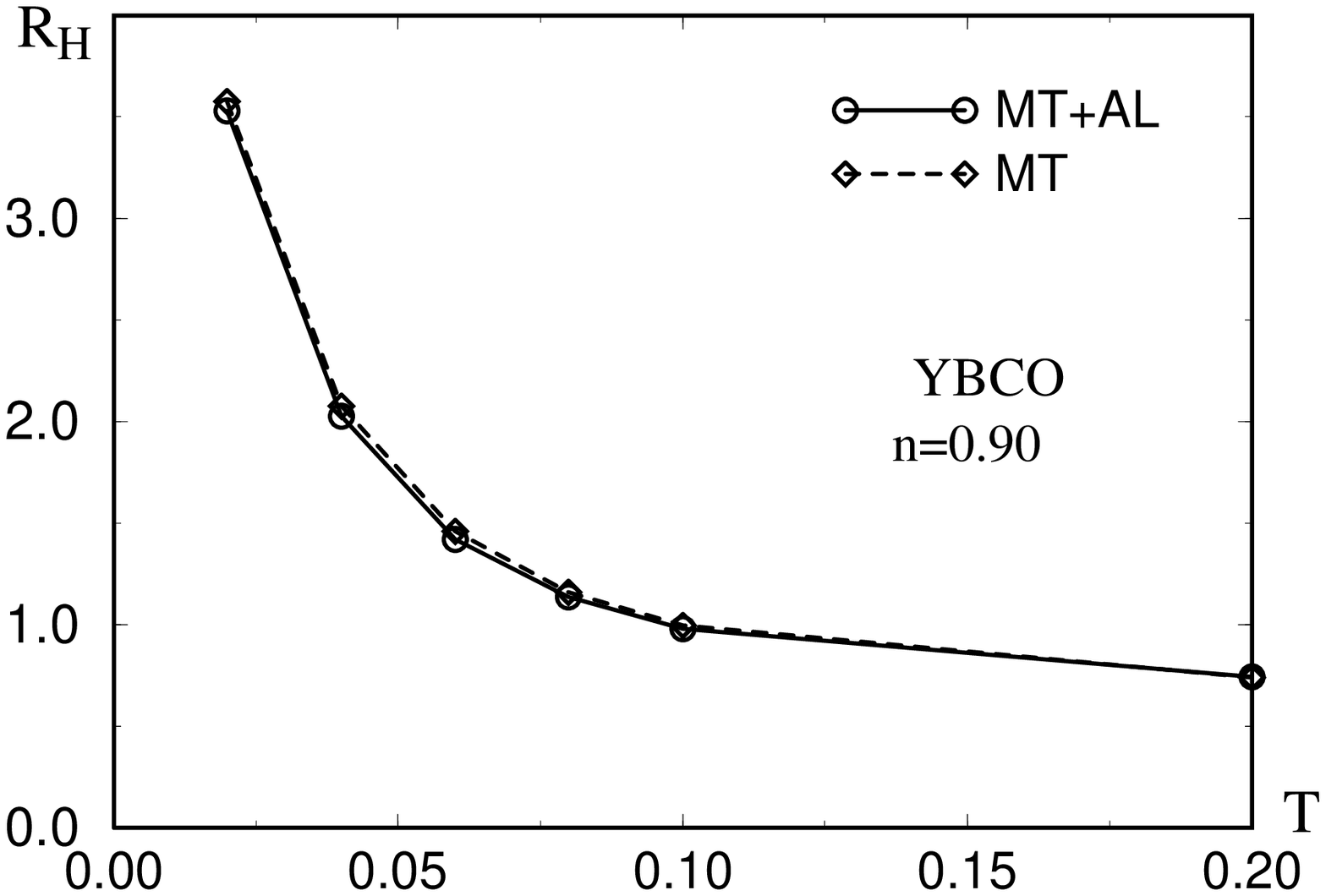}}
\caption{}
 The obtained $T$-dependence of the resistivity and the Hall
 coefficient. MT and MT+AL are given by $J_{\k\mu}(\w)$ 
 derived from
eq. (\ref{eqn:J_numerical}) and (\ref{eqn:J_numerical_all}), 
 respectively.
\label{fig:Hall2}
\end{figure}

Figure \ref{fig:Hall2} shows the calculated resistivity and the
Hall coefficient for YBCO $(n=0.90)$
derived from $J_{\k \mu}^{\rm MT+AL}(\w)$, 
together with those given in \S VI B.
We see that the AL terms give only small corrections
to $\rho$ and $R_{\rm H}$.

\end{multicols}

\begin{references}
\bibitem{pseudo_scaling}
 The ARPES and STM experiments show that
 the pseudo gap appears in the density of states
 below $T^{\ast\ast}$ $(>T^{\ast})$.
 Experimentally, $\w_{\rm sf}\cdot \xi$ is almost constant
 and $\xi\propto 1/T$ for $T^{\ast\ast}>T>T^\ast$.
 However, the crossover phenomena at $T=T^{\ast\ast}$ 
 are rather moderate and obscure.
\end{references}
\end{document}